\documentstyle[10pt,epsf,epsfig,dp_delphititle]{dp_delphi}
\makeindex
\def\DpPaperGroup{PH--EP}
\def\DpPaperRef{2008-018}
\def\DpDate{13 November 2008}
\def\DpAuthors{DELPHI Collaboration}
\def\DpSubmit{(Accepted by Eur. Phys. J. C)}
\def\DpTitle{{
A Study of \bbbar\ Production in \ee\ Collisions at $\sqrt{s} =$ 130-207 GeV
}}
\def\DpComment{}
\def\DpEMail{}

\newcommand{\ee}{\mbox{${\mathrm e}^+{\mathrm e}^-$}}

\newcommand{\qqbar}{\mbox{${\mathrm q}\bar{\mathrm q}$}}
\newcommand{\bbbar}{\mbox{${\mathrm b}\bar{\mathrm b}$}}
\newcommand{\ccbar}{\mbox{${\mathrm c}\bar{\mathrm c}$}}

\newcommand{\Rb} {\mbox{$R_{\rm b}$}}
\newcommand{\Ab} {\mbox{$A^{\rm b}_{\rm FB}$}}
\newcommand{\Bt} {\mbox{$B_{\rm tag}$}}

\begin{document}
%%%%%%%%%%%%%%%%%%%%%%%%%% Is there a problem with Coll.Sty ?
\makeatletter
%\input{dp_system:coll.sty}
% Collapse citation numbers to ranges.  Non-numeric and undefined labels
% are handled.  No sorting is done.  E.g., 1,3,2,3,4,5,foo,1,2,3,?,4,5
% gives 1,3,2-5,foo,1-3,?,4,5
\newcount\@tempcntc
\def\@citex[#1]#2{\if@filesw\immediate\write\@auxout{\string\citation{#2}}\fi
  \@tempcnta\z@\@tempcntb\m@ne\def\@citea{}\@cite{\@for\@citeb:=#2\do
    {\@ifundefined
       {b@\@citeb}{\@citeo\@tempcntb\m@ne\@citea\def\@citea{,}{\bf ?}\@warning
       {Citation `\@citeb' on page \thepage \space undefined}}%
    {\setbox\z@\hbox{\global\@tempcntc0\csname b@\@citeb\endcsname\relax}%
     \ifnum\@tempcntc=\z@ \@citeo\@tempcntb\m@ne
       \@citea\def\@citea{,}\hbox{\csname b@\@citeb\endcsname}%
     \else
      \advance\@tempcntb\@ne
      \ifnum\@tempcntb=\@tempcntc
      \else\advance\@tempcntb\m@ne\@citeo
      \@tempcnta\@tempcntc\@tempcntb\@tempcntc\fi\fi}}\@citeo}{#1}}
\def\@citeo{\ifnum\@tempcnta>\@tempcntb\else\@citea\def\@citea{,}%
  \ifnum\@tempcnta=\@tempcntb\the\@tempcnta\else
   {\advance\@tempcnta\@ne\ifnum\@tempcnta=\@tempcntb \else \def\@citea{--}\fi
    \advance\@tempcnta\m@ne\the\@tempcnta\@citea\the\@tempcntb}\fi\fi}
 
\makeatother
%%%%%%%%%%%%%%%%%%%%%%%%%% ??????????????????????????????????
% Generate the title page
\begin{titlepage}
\pagenumbering{roman}
\CERNpreprint{\DpPaperGroup}{\DpPaperRef} % Reference of the paper
\date{{\small\DpDate}} % Date of the paper
\title{\DpTitle} % Title of the paper
\address{\DpAuthors} % General name of the author(s)

\begin{shortabs} % Start the abstract
\noindent
Measurements are presented of \Rb, the ratio of the \bbbar\ cross-section
to the \qqbar\ cross-section in \ee\ collisions, and the forward-backward 
asymmetry \Ab\ at twelve energy points in the range $\sqrt{s} = 130-207$~GeV.
These results are found to be consistent with the Standard Model expectations.
The measurements are used to set limits on new physics scenarios involving
contact interactions.

\end{shortabs}

\vfill

\begin{center}
\DpSubmit \ \\ % Horrible hack to allow to have DpSubmit empty
\DpComment \ \\
\DpEMail \ \\
\end{center}
\vfill
\clearpage

\headsep 10.0pt

\addtolength{\textheight}{10mm}
\addtolength{\footskip}{-5mm}
\begingroup
% Commands to process the author names
%
\newcommand{\DpName}[2]{\hbox{#1$^{\ref{#2}}$},\hfill}
\newcommand{\DpNameTwo}[3]{\hbox{#1$^{\ref{#2},\ref{#3}}$},\hfill}
\newcommand{\DpNameThree}[4]{\hbox{#1$^{\ref{#2},\ref{#3},\ref{#4}}$},\hfill}
\newskip\Bigfill \Bigfill = 0pt plus 1000fill
\newcommand{\DpNameLast}[2]{\hbox{#1$^{\ref{#2}}$}\hspace{\Bigfill}}
%
%\small
\footnotesize
\noindent
\DpName{J.Abdallah}{LPNHE}
\DpName{P.Abreu}{LIP}
\DpName{W.Adam}{VIENNA}
\DpName{P.Adzic}{DEMOKRITOS}
\DpName{T.Albrecht}{KARLSRUHE}
\DpName{R.Alemany-Fernandez}{CERN}
\DpName{T.Allmendinger}{KARLSRUHE}
\DpName{P.P.Allport}{LIVERPOOL}
\DpName{U.Amaldi}{MILANO2}
\DpName{N.Amapane}{TORINO}
% ...
\normalsize
\endgroup
\newpage

\titlefoot{Department of Physics and Astronomy, Iowa State
     University, Ames IA 50011-3160, USA
    \label{AMES}}
\titlefoot{Physics Department, Universiteit Antwerpen,
     Universiteitsplein 1, B-2610 Antwerpen, Belgium
    \label{ANTWERP}}
\titlefoot{IIHE, ULB-VUB,
     Pleinlaan 2, B-1050 Brussels, Belgium
    \label{BRUSSELS}}
\titlefoot{Physics Laboratory, University of Athens, Solonos Str.
     104, GR-10680 Athens, Greece
    \label{ATHENS}}
\titlefoot{Department of Physics, University of Bergen,
     All\'egaten 55, NO-5007 Bergen, Norway
    \label{BERGEN}}
\titlefoot{Dipartimento di Fisica, Universit\`a di Bologna and INFN,
     Via Irnerio 46, IT-40126 Bologna, Italy
    \label{BOLOGNA}}
\titlefoot{Centro Brasileiro de Pesquisas F\'{\i}sicas, rua Xavier Sigaud 150,
     BR-22290 Rio de Janeiro, Brazil
    \label{BRASIL-CBPF}}
\titlefoot{Inst. de F\'{\i}sica, Univ. Estadual do Rio de Janeiro,
     rua S\~{a}o Francisco Xavier 524, Rio de Janeiro, Brazil
    \label{BRASIL-IFUERJ}}
\titlefoot{Coll\`ege de France, Lab. de Physique Corpusculaire, IN2P3-CNRS,
     FR-75231 Paris Cedex 05, France
    \label{CDF}}
\titlefoot{CERN, CH-1211 Geneva 23, Switzerland
    \label{CERN}}
\titlefoot{Institut de Recherches Subatomiques, IN2P3 - CNRS/ULP - BP20,
     FR-67037 Strasbourg Cedex, France
    \label{CRN}}
\titlefoot{Now at DESY-Zeuthen, Platanenallee 6, D-15735 Zeuthen, Germany
    \label{DESY}}
\titlefoot{Institute of Nuclear Physics, N.C.S.R. Demokritos,
     P.O. Box 60228, GR-15310 Athens, Greece
    \label{DEMOKRITOS}}
\titlefoot{FZU, Inst. of Phys. of the C.A.S. High Energy Physics Division,
     Na Slovance 2, CZ-182 21, Praha 8, Czech Republic
    \label{FZU}}
\titlefoot{Dipartimento di Fisica, Universit\`a di Genova and INFN,
     Via Dodecaneso 33, IT-16146 Genova, Italy
    \label{GENOVA}}
\titlefoot{Institut des Sciences Nucl\'eaires, IN2P3-CNRS, Universit\'e
     de Grenoble 1, FR-38026 Grenoble Cedex, France
    \label{GRENOBLE}}
\titlefoot{Helsinki Institute of Physics and Department of Physical Sciences,
     P.O. Box 64, FIN-00014 University of Helsinki, 
     \indent~~Finland
    \label{HELSINKI}}
\titlefoot{Joint Institute for Nuclear Research, Dubna, Head Post
     Office, P.O. Box 79, RU-101 000 Moscow, Russian Federation
    \label{JINR}}
\titlefoot{Institut f\"ur Experimentelle Kernphysik,
     Universit\"at Karlsruhe, Postfach 6980, DE-76128 Karlsruhe,
     Germany
    \label{KARLSRUHE}}
\titlefoot{Institute of Nuclear Physics PAN,Ul. Radzikowskiego 152,
     PL-31142 Krakow, Poland
    \label{KRAKOW1}}
\titlefoot{Faculty of Physics and Nuclear Techniques, University of Mining
     and Metallurgy, PL-30055 Krakow, Poland
    \label{KRAKOW2}}
%\titlefoot{Universit\'e de Paris-Sud, Lab. de l'Acc\'el\'erateur
%    Lin\'eaire, IN2P3-CNRS, B\^{a}t. 200, FR-91405 Orsay Cedex, France
\titlefoot{LAL, Univ Paris-Sud, CNRS/IN2P3, Orsay, France
    \label{LAL}}
\titlefoot{School of Physics and Chemistry, University of Lancaster,
     Lancaster LA1 4YB, UK
    \label{LANCASTER}}
\titlefoot{LIP, IST, FCUL - Av. Elias Garcia, 14-$1^{o}$,
     PT-1000 Lisboa Codex, Portugal
    \label{LIP}}
\titlefoot{Department of Physics, University of Liverpool, P.O.
     Box 147, Liverpool L69 3BX, UK
    \label{LIVERPOOL}}
\titlefoot{Dept. of Physics and Astronomy, Kelvin Building,
     University of Glasgow, Glasgow G12 8QQ, UK
    \label{GLASGOW}}
\titlefoot{LPNHE, IN2P3-CNRS, Univ.~Paris VI et VII, Tour 33 (RdC),
     4 place Jussieu, FR-75252 Paris Cedex 05, France
    \label{LPNHE}}
\titlefoot{Department of Physics, University of Lund,
     S\"olvegatan 14, SE-223 63 Lund, Sweden
    \label{LUND}}
\titlefoot{Universit\'e Claude Bernard de Lyon, IPNL, IN2P3-CNRS,
     FR-69622 Villeurbanne Cedex, France
    \label{LYON}}
\titlefoot{Dipartimento di Fisica, Universit\`a di Milano and INFN-MILANO,
     Via Celoria 16, IT-20133 Milan, Italy
    \label{MILANO}}
\titlefoot{Dipartimento di Fisica, Univ. di Milano-Bicocca and
     INFN-MILANO, Piazza della Scienza 3, IT-20126 Milan, Italy
    \label{MILANO2}}
\titlefoot{IPNP of MFF, Charles Univ., Areal MFF,
     V Holesovickach 2, CZ-180 00, Praha 8, Czech Republic
    \label{NC}}
\titlefoot{NIKHEF, Postbus 41882, NL-1009 DB
     Amsterdam, The Netherlands
    \label{NIKHEF}}
\titlefoot{National Technical University, Physics Department,
     Zografou Campus, GR-15773 Athens, Greece
    \label{NTU-ATHENS}}
\titlefoot{Physics Department, University of Oslo, Blindern,
     NO-0316 Oslo, Norway
    \label{OSLO}}
\titlefoot{Dpto. Fisica, Univ. Oviedo, Avda. Calvo Sotelo
     s/n, ES-33007 Oviedo, Spain
    \label{OVIEDO}}
\titlefoot{Department of Physics, University of Oxford,
     Keble Road, Oxford OX1 3RH, UK
    \label{OXFORD}}
\titlefoot{Dipartimento di Fisica, Universit\`a di Padova and
     INFN, Via Marzolo 8, IT-35131 Padua, Italy
    \label{PADOVA}}
\titlefoot{Rutherford Appleton Laboratory, Chilton, Didcot
     OX11 OQX, UK
    \label{RAL}}
\titlefoot{Dipartimento di Fisica, Universit\`a di Roma II and
     INFN, Tor Vergata, IT-00173 Rome, Italy
    \label{ROMA2}}
\titlefoot{Dipartimento di Fisica, Universit\`a di Roma III and
     INFN, Via della Vasca Navale 84, IT-00146 Rome, Italy
    \label{ROMA3}}
\titlefoot{DAPNIA/Service de Physique des Particules,
     CEA-Saclay, FR-91191 Gif-sur-Yvette Cedex, France
    \label{SACLAY}}
\titlefoot{Instituto de Fisica de Cantabria (CSIC-UC), Avda.
     los Castros s/n, ES-39006 Santander, Spain
    \label{SANTANDER}}
\titlefoot{Inst. for High Energy Physics, Serpukov
     P.O. Box 35, Protvino, (Moscow Region), Russian Federation
    \label{SERPUKHOV}}
\titlefoot{J. Stefan Institute, Jamova 39, SI-1000 Ljubljana, Slovenia
    \label{SLOVENIJA1}}
\titlefoot{Laboratory for Astroparticle Physics,
     University of Nova Gorica, Kostanjeviska 16a, SI-5000 Nova Gorica, Slovenia
    \label{SLOVENIJA2}}
\titlefoot{Department of Physics, University of Ljubljana,
     SI-1000 Ljubljana, Slovenia
    \label{SLOVENIJA3}}
\titlefoot{Fysikum, Stockholm University,
     Box 6730, SE-113 85 Stockholm, Sweden
    \label{STOCKHOLM}}
\titlefoot{Dipartimento di Fisica Sperimentale, Universit\`a di
     Torino and INFN, Via P. Giuria 1, IT-10125 Turin, Italy
    \label{TORINO}}
\titlefoot{INFN,Sezione di Torino and Dipartimento di Fisica Teorica,
     Universit\`a di Torino, Via Giuria 1,
     IT-10125 Turin, Italy
    \label{TORINOTH}}
\titlefoot{Dipartimento di Fisica, Universit\`a di Trieste and
     INFN, Via A. Valerio 2, IT-34127 Trieste, Italy
    \label{TRIESTE}}
\titlefoot{Istituto di Fisica, Universit\`a di Udine and INFN,
     IT-33100 Udine, Italy
    \label{UDINE}}
\titlefoot{Univ. Federal do Rio de Janeiro, C.P. 68528
     Cidade Univ., Ilha do Fund\~ao
     BR-21945-970 Rio de Janeiro, Brazil
    \label{UFRJ}}
\titlefoot{Department of Radiation Sciences, University of
     Uppsala, P.O. Box 535, SE-751 21 Uppsala, Sweden
    \label{UPPSALA}}
\titlefoot{IFIC, Valencia-CSIC, and D.F.A.M.N., U. de Valencia,
     Avda. Dr. Moliner 50, ES-46100 Burjassot (Valencia), Spain
    \label{VALENCIA}}
\titlefoot{Institut f\"ur Hochenergiephysik, \"Osterr. Akad.
     d. Wissensch., Nikolsdorfergasse 18, AT-1050 Vienna, Austria
    \label{VIENNA}}
\titlefoot{Inst. Nuclear Studies and University of Warsaw, Ul.
     Hoza 69, PL-00681 Warsaw, Poland
    \label{WARSZAWA}}
\titlefoot{Now at University of Warwick, Coventry CV4 7AL, UK
    \label{WARWICK}}
\titlefoot{Fachbereich Physik, University of Wuppertal, Postfach
     100 127, DE-42097 Wuppertal, Germany \\
\noindent
{$^\dagger$~deceased}
    \label{WUPPERTAL}}
\addtolength{\textheight}{-10mm}
\addtolength{\footskip}{5mm}
\clearpage
\headsep 30.0pt
\end{titlepage}
%%%%%%%%%%%%%%%%%%%%%%%%%
%
% Change for the document body
%%\pagestyle{heading} % for page numbering
\pagenumbering{arabic} % page numbering in number
\setcounter{footnote}{0} %
\large
%\linenumbers %%%CD

%\input{document.tex}    % The body of the document.
\section{Introduction}

The ratio $R_{\rm b} \equiv \sigma(\ee \to \bbbar)/\sigma(\ee \to \qqbar)$
%~\footnote{
%Note that in LEP~1 analyses the alternative definition  
%$R_{\rm b} \equiv \Gamma({\rm Z} \to \bbbar)/\sigma({\rm Z} \to \qqbar)$ was employed.}
and \Ab, the forward-backward production asymmetry of bottom quarks
in \ee\ collisions,
are important parameters in precision studies of electroweak theory,
and are sensitive probes of new physics.  
%
%In the Standard Model $\ee \to \bbbar$ events are produced by an s-channel
%process propagated by either a photon or Z-boson exchange.   
%
This paper presents
measurements of \Rb\ and \Ab\ made 
at centre-of-mass energies ($\sqrt{s}$) between 130~GeV and 207~GeV.
%
%In this regime the relative contributions of the photon and Z-propagators 
%are such that the value of $R_{\rm b}$ is expected to fall slowly,
%and that of \Ab\ to rise, with collision energy.
%
Events containing a \bbbar\ pair have several characteristic
features, most notably the presence of secondary vertices,
which may be used to select a sample enriched in b-decays.
A `b-tag' variable has been constructed for this purpose,
which exploits the high resolution tracking provided by 
the DELPHI Silicon Tracker.
In the asymmetry measurement the hemisphere containing the b-quark has been
determined using a hemisphere-charge technique.   In order to enhance sensitivity
to possible new physics contributions from high energy scales, all measurements 
have been made for events in which $\sqrt{s'/s} \ge 0.85$, where $\sqrt{s'}$ is
the effective centre-of-mass energy after initial state radiation.
In the Standard Model $\ee \to \bbbar$ events are produced by an s-channel
process propagated by either photon or Z-boson exchange.
Over the interval of collision energies under investigation
the relative strengths of the two contributions evolve so 
that the value of $R_{\rm b}$ is expected to fall,
and that of \Ab\ to rise, slowly with $\sqrt{s}$.

Studies of \bbbar\ production at collision energies above the Z-pole have
been presented by other LEP collaborations~\cite{ALEPHBB1,ALEPHBB2,L3BB,OPALBB1,OPALBB2}.  The results presented here
for the energies $130 \le \sqrt{s} \le 172$~GeV supersede those of 
an earlier DELPHI publication~\cite{DELPHIOLD}. 

Sect.~\ref{sec:dataset} describes the datasets and the aspects of the DELPHI
detector relevant for the analysis.  The event selection is discussed in
Sect.~\ref{sec:selection}.  The \Rb\ determination is presented in Sect.~\ref{sec:rb}
and that of \Ab\ in Sect.~\ref{sec:afb}.   An interpretation of the results
within the context of both the Standard Model and possible new physics
models including contact interactions is given in Sect.~\ref{sec:interpret}.

\section{Datasets, the DELPHI Detector and Simulation}
\label{sec:dataset}

LEP~2 operation began in 1995, when around 6~$\rm pb^{-1}$ of data were delivered at centre-of-mass energies
of $\sqrt{s}=$130~GeV and 136~GeV.  In 1996 the collision energy of the beams
was raised to, and then beyond, the $\rm W^+W^-$ production threshold of 161~GeV.  Each subsequent year
saw increasing amounts of integrated luminosity produced at ever higher energies,
reaching 209~GeV in the year 2000.   In total around 680~$\rm pb^{-1}$ were collected by 
the DELPHI experiment at 12 separate energy points.     
Note that during the 2000 run, operation occurred at a near-continuum
of energies between 202~GeV and 209~GeV.  In the present study the data collected during 2000 
are divided into two bins, above and below 205.5~GeV.   
Throughout LEP~2 operation collisions were performed with unpolarised beams.
The mean collision energies for each 
period of operation and
the integrated luminosities used in the analysis  are summarised in  Table~\ref{tab:evtall}.
More details on the LEP collision energy calibration and the DELPHI luminosity determination
are given in~\cite{ECAL} and~\cite{DELPHIFF}, respectively.

In addition to the high energy operation, in each year from 1996 onwards 
LEP also delivered 1--4 $\rm pb^{-1}$ 
at the Z-pole,
in order to provide well understood calibration data for the experiments.
In this paper the events collected during the calibration running 
are referred to as the `Z-data', and provide control samples for the high-energy studies.
In 1995 the control sample is taken from the Z-peak data immediately preceeding the
switch to 130~GeV operation.  In 2000 a second set of Z-data was collected
in order to provide a dedicated calibration sample for the period in which
the DELPHI TPC had impaired efficiency (see below).

A description of the DELPHI detector and its performance can be found in~\cite{DELPHIDET,DELPHIDET2}.
For the analyses presented in this paper, 
the most important sub-detector in DELPHI was the Silicon Tracker~\cite{ST}.
The Silicon Tracker was a three-layer vertex detector providing measurements in
both the views transverse and longitudinal to the beam line, 
with the capabilities to provide effective b-tagging over the polar angle interval of 
$25^\circ <\theta < 155^\circ$, where $\theta$ is the angle with respect to the $\rm e^-$ beam
direction.  End-caps
of mini-strip and pixel detectors gave tracking coverage down to $\theta = 10^\circ \, (170^\circ)$.
The Silicon Tracker was fully installed in 1996 and remained operational until the end of the
LEP~2 programme.   During the 1995 run b-tagging information was
provided by the microvertex detector described in~\cite{VD}.

During the 2000 run, one of the 12 azimuthal sectors of the central tracking chamber, the TPC, failed.
After the beginning of September 2000 it was not possible to detect the tracks
left by charged particles in that sector.  The data affected correspond to approximately one quarter
of the total dataset of that year (the `BTPC' period).  
Nevertheless, the redundancy of the tracking system of 
DELPHI meant that tracks passing through the sector could still be reconstructed from signals
in the other tracking detectors.  A modified tracking reconstruction algorithm was used in this
sector, which included space points reconstructed in the Barrel RICH detector.  As a result,
the track reconstruction efficiency was only slightly reduced in the region covered by the
broken sector, but the track parameter resolutions were degraded compared with the data taken
prior to the failure of this sector (the `GTPC' period).
% For the studies reported here,  the 
%GTPC and BTPC data were analysed separately, making use of detector simulation appropriate for 
%each period.

To determine selection efficiencies and backgrounds in the analysis, events were simulated
using a variety of generators and the DELPHI Monte Carlo~\cite{DELPHIDET2}. These events were
passed through the full data analysis chain.  Different software versions were
used for each year, in order to follow time variations in the detector performance.
For the year 2000, separate GTPC and BTPC sets of simulation were produced.
The typical size of the simulated samples used in the analysis is two orders of magnitude larger
than those of the data.

The $\rm e^+e^- \to f\overline{f}$ process was simulated with KK 4.14~\cite{KK},
interfaced with PYTHIA 6.156~\cite{PYTHIA1,PYTHIA2} for the description of the hadronisation.
For systematic studies, the alternative hadronisation description implemented
in ARIADNE 4.08~\cite{ARIADNE} was used.   Four-fermion background events
were simulated with the generator WPHACT 2.0~\cite{WPHACT1,WPHACT2}, with PYTHIA again
used for the hadronisation.

\section{Event Selection}
\label{sec:selection}

The analysis was made using charged particles with momentum lying between 0.1~GeV and 1.5$\cdot (\sqrt{s}/2)$,
and measurement uncertainty of less than 100\%, and having a closest approach to the beam-spot of less
than 4~cm in the plane perpendicular to the beam axis, and less than 4/$\sin \theta$~cm along the beam axis.
Neutral showers were used above a minimum energy cut, which was 300~MeV for the barrel electromagnetic (HPC)
and very forward calorimeter (STIC), and 400~MeV for the forward electromagnetic calorimeter (FEMC).

The following requirements were applied to select a pure sample of hadronic events,
and to ensure that each event lay within the acceptance of the Silicon Tracker:
\begin{itemize}
\item{Number of charged particle tracks $\ge$ 7;}
\item{Quadrature sum over each end-cap 
of energy reconstructed in the forward electromagnetic calorimeter system (STIC + FEMC)
$\le 0.85 (\sqrt{s}/2)$;} 
\item{Total transverse energy $> 0.2 \sqrt{s}$;}
\item{Energy of charged particles $> 0.1 \sqrt{s}$;}
\item{Restriction on the polar angle of the thrust axis, $\theta_T$, such that $|\cos \theta_T| \le 0.9$.}
\end{itemize}
Data-taking runs were excluded in which the tracking detectors and Silicon Tracker were not fully operational.

In addition to this selection a `W-veto' was applied to suppress the contamination from four-fermion
events.  The veto procedure consisted of forcing the 
event into a four-jet topology using the LUCLUS~\cite{PYTHIA1,PYTHIA2} algorithm and imposing the requirement that
$(E_{\rm min}/\sqrt{s}) \cdot \alpha_{\rm min}\, <\, 4.25^\circ$, where $E_{\rm min}$ is the energy 
of the
softest jet, and $\alpha_{\rm min}$ the smallest opening angle found between all two-jet combinations.
This condition is designed to distinguish between two-fermion events containing gluon jets, and
genuine four-fermion background.  Less than 40\% of four-fermion events survive the hadronic
selection and the W-veto.

%After this selection the contamination from $\ee \to l^+l^-$ events and $\gamma \gamma$ collisions
%is negligible.  The background from four-fermion events rises from approximately 9\% in the 172.1~GeV
%dataset, to 21\% in the 206.6~GeV sample. 

The analysis is concerned with events produced with an 
effective centre-of-mass energy of the $\rm q\overline{q}$ system, $\sqrt{s'}$, 
at or around the collision
energy, $\sqrt{s}$. The effective  centre-of-mass energy 
is reconstructed as in the hadronic analysis reported in~\cite{DELPHIFF}.
A constrained fit is performed, taking as input the observed jet directions 
as found by the DURHAM clustering algorithm~\cite{DURHAM}, imposing
energy and momentum conservation, and assuming any ISR photon was emitted along the beam line.
Radiative returns to the Z are then rejected by requiring that the reconstructed value of 
$\sqrt{s'/s} \ge 0.85.$
Contamination from events with true values of $\sqrt{s'/s}$ below this threshold
is around 16\% at 130.3~GeV and reduces to about 6\% at 206.6~GeV.

As a final condition, events with $|Q^+_{\rm FB}|\ge 1.5$ are rejected, where
$|Q^+_{\rm FB}|$ is one of the event charge variables
defined in Sect.~\ref{sec:afbproc}.   This selection is applied to exclude
badly measured events from the asymmetry measurement, and removes around
0.5\% of the sample. 

The numbers of events passing the high $\sqrt{s'/s}$ two-fermion hadronic  selection 
at each energy point are listed in Table~\ref{tab:evtall}, together with the Monte Carlo expectations.
The two sets of numbers agree well.  The background from four-fermion events is estimated
to be around 9\% in the 172.1~GeV dataset, rising to 21\% in the 206.6~GeV sample.
The contamination from $\tau^+\tau^-$ events is around 0.3\%.  All other backgrounds 
are negligible.

\begin{table}
\begin{center}
\caption[]{The year of data-taking, mean centre-of-mass energy, 
integrated luminosity, number of events after hadronic selection and W-rejection (`Before b-tag'),  
and number of events after the b-tag.  In the year 2000 the numbers in parentheses are those corresponding to
the GTPC sub-sample. Numbers are shown for data and Monte Carlo,
where for the latter the samples have been scaled to the integrated luminosity of the data and  
Standard Model cross-section values are assumed.}
\label{tab:evtall}
\vspace*{0.2cm}
\begin{tabular}{c|c|r|rr|rr}
     &             &                             & \multicolumn{2}{|c|}{\hspace{1.0cm}Before b-tag} & \multicolumn{2}{c}{\hspace{0.5cm} After b-tag} \\
Year & $\sqrt{s}$ [GeV] &  $\int {\cal L} \, \rm dt$ [$\rm pb^{-1}$] &  Data &  MC  \hspace{0.0cm}        &   Data   &  \hspace{0.0cm} MC  \hspace{0.0cm}          \\ \hline
1995 &130.3            &     2.9   \hspace{0.6cm}        &     224 \hspace{0.0cm}         &   224 \hspace{0.0cm}   &      30 \hspace{0.0cm} &  24   \hspace{0.1cm}\\
     &136.3            &     2.6   \hspace{0.6cm}        &     160 \hspace{0.0cm}         &   160 \hspace{0.0cm}   &      15 \hspace{0.0cm} &  17   \hspace{0.1cm}\\ \hline
1996 &161.3            &    10.1   \hspace{0.6cm}        &     363 \hspace{0.0cm}         &   321 \hspace{0.0cm}   &      46 \hspace{0.0cm} &  36   \hspace{0.1cm}\\
     &172.1            &    10.0   \hspace{0.6cm}        &     304 \hspace{0.0cm}         &   280 \hspace{0.0cm}   &      27 \hspace{0.0cm} &  29   \hspace{0.1cm}\\ \hline
1997 &182.7            &    53.1   \hspace{0.6cm}        &    1351 \hspace{0.0cm}         &  1284 \hspace{0.0cm}   &     117 \hspace{0.0cm} & \hspace{0.1cm} 137   \hspace{0.0cm} \\ \hline
1998 &188.6            &   156.8   \hspace{0.6cm}        &    3567 \hspace{0.0cm}         &  3541 \hspace{0.0cm}   &     365 \hspace{0.0cm} & 379   \hspace{0.1cm}\\ \hline
1999 &191.6            &    25.8   \hspace{0.6cm}        &     563 \hspace{0.0cm}         &   565 \hspace{0.0cm}   &      68 \hspace{0.0cm} &  57   \hspace{0.1cm}\\ 
     &195.5            &    76.2   \hspace{0.6cm}        &    1629 \hspace{0.0cm}         &  1597 \hspace{0.0cm}   &     164 \hspace{0.0cm} & 159   \hspace{0.1cm}\\
     &199.5            &    83.0   \hspace{0.6cm}        &    1651 \hspace{0.0cm}         &  1670 \hspace{0.0cm}   &     184 \hspace{0.0cm} & 162   \hspace{0.1cm}\\ 
     &201.7            &    40.6   \hspace{0.6cm}        &     807 \hspace{0.0cm}         &   799 \hspace{0.0cm}   &      88 \hspace{0.0cm} &  77   \hspace{0.1cm}\\ \hline
2000 &204.8 (204.8)    &    82.8  (76.1)                 &    1538  (1411)                &  1572   (1447)         &     144   (131)        & 147    (137)        \\
     &206.6 (206.6)    &   136.4  (84.7)                 &    2510  (1586)                &  2536   (1581)         &     240   (167)        & 233    (148)        \\ \hline
     &Total            &   680.3   \hspace{0.6cm}        &   14667 \hspace{0.0cm}         & 14549 \hspace{0.0cm}   &    1488 \hspace{0.0cm} &\hspace{0.1cm} 1457   \hspace{0.0cm} \\
\end{tabular}
\end{center}
\end{table}

A `b-tag' variable is used to extract a sub-sample of events enriched 
in b-quarks from the non-radiative $\rm q \overline{q}$ sample.
This variable makes use of three observables, known to distinguish
between b-quark events and those events with non-b content.
In this analysis, the three categories of observable considered are:
\begin{itemize}
\item{A lifetime variable, constructed from the impact parameters of
charged particle tracks in each jet;}
\item{The invariant mass of charged particles forming any secondary
vertices that are found;}
\item{The rapidities of charged particles in any secondary vertex, defined
with respect to the jet direction.}
\end{itemize}
These properties are used to construct a single event `b-tag' variable,
\Bt, of typical value between -5 and 10.  Events with higher values of this
variable are enriched in b-events. 
More information on the b-tagging procedure may be 
found in~\cite{BTAG}.   
In this analysis a cut
value of 1 is used for all high energy data sets to select the b-enriched sample; this selection
has a typical efficiency for \bbbar\ events of around 65\%,
but only 2.5\% for \ccbar\ events and 0.3\% for light quark events.  
The numbers of events passing the b-tag 
are listed in Table~\ref{tab:evtall}.  Here the Monte Carlo
numbers 
%assume the Standard Model values of \Rb\ and include 
do not include the correction factors discussed in Sect.~\ref{sec:rb}.

\section{Measurement of \Rb}
\label{sec:rb}

\subsection{Procedure and Calibration with Z Data}

For each energy point \Rb\ is determined through the following relation:
\begin{equation}
\frac{N_{\rm tag}^{\rm D} \, - \, N_{\rm tag}^{\rm 4f}}
     {N_{\rm total}^{\rm D} \, - \, N_{\rm total}^{\rm 4f}} \: = \:
\Rb \,c_b\, \epsilon_b \: + \: R_c \, c_c \, \epsilon_c \: + \:
\epsilon_{uds} \, \left ( 1 \, - \, c_c \, R_c \, - \, c_b \, R_b \right ).
\label{eq:rbextract}
\end{equation}
Here $N_{\rm total \, (tag)}^{\rm D}$ and $N_{\rm total \, (tag)}^{\rm 4f}$ 
are the number of events in the
data, and the estimated four-fermion background respectively, before (after)
the application of the b-tag cut; $R_{\rm c}$ is directly analogous to \Rb, 
but defined for $\rm c\bar{c}$ events; and $\epsilon_b$, $\epsilon_c$ and 
$\epsilon_{uds}$ are the efficiencies
of the b-tag cut applied to b, c and light quark events respectively. 
$c_b$ and $c_c$ are correction factors,  which account for the fact
that the effective values of \Rb\ and $R_{\rm c}$ are modified by the
hadronic selection, and that there is some contamination from initial state 
radiative production 
in the sample, the fraction of which can in principle be different
for each quark type, and therefore changes with the application of the b-tag.
Simulation indicated that these correction factors lie within 1-2\% of 
unity.

\begin{figure}
\vspace*{-1.2cm}
\begin{center}
\epsfig{file=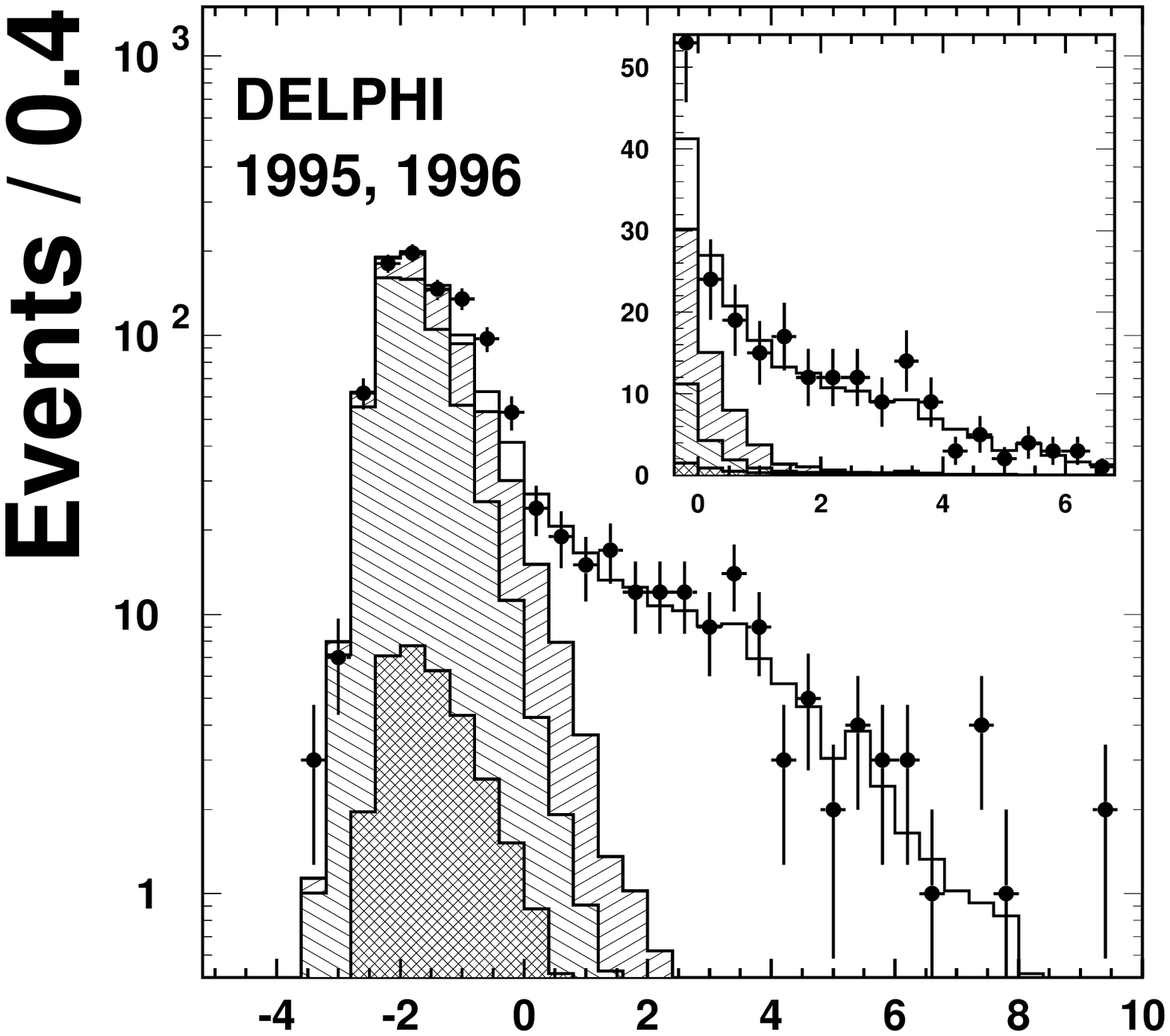,width=0.48\textwidth}
\epsfig{file=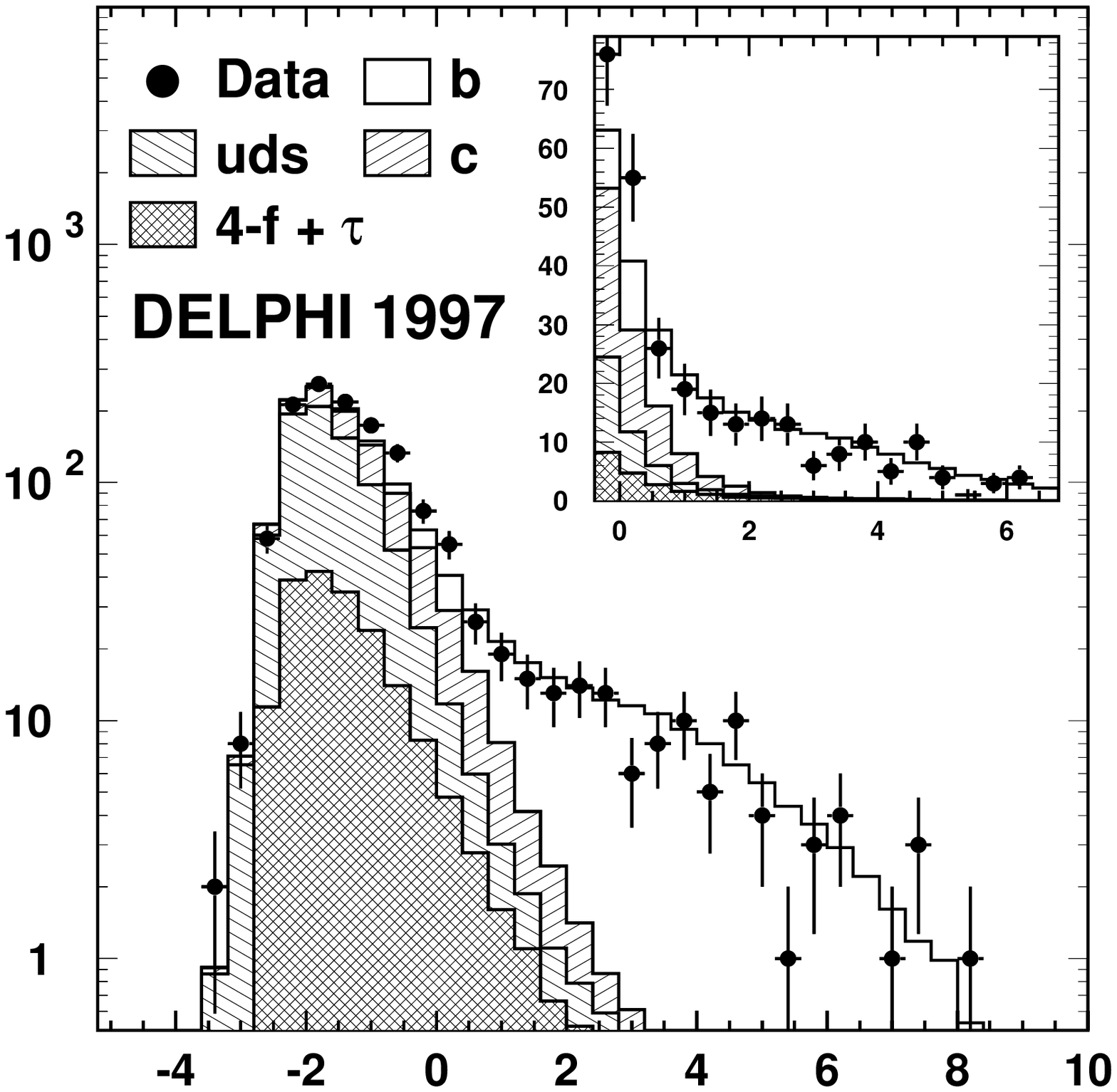,width=0.48\textwidth}
\epsfig{file=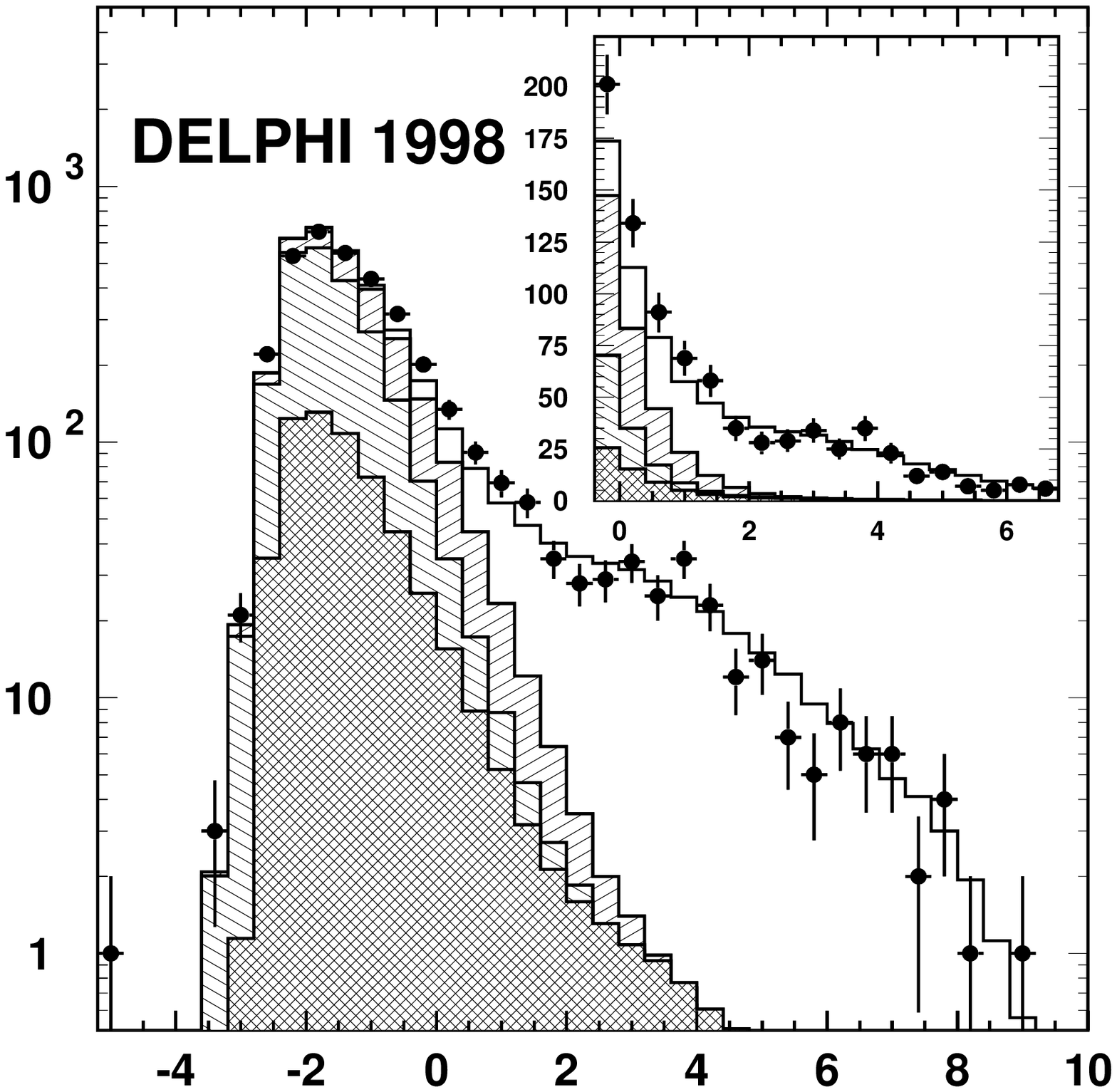,width=0.48\textwidth}
\epsfig{file=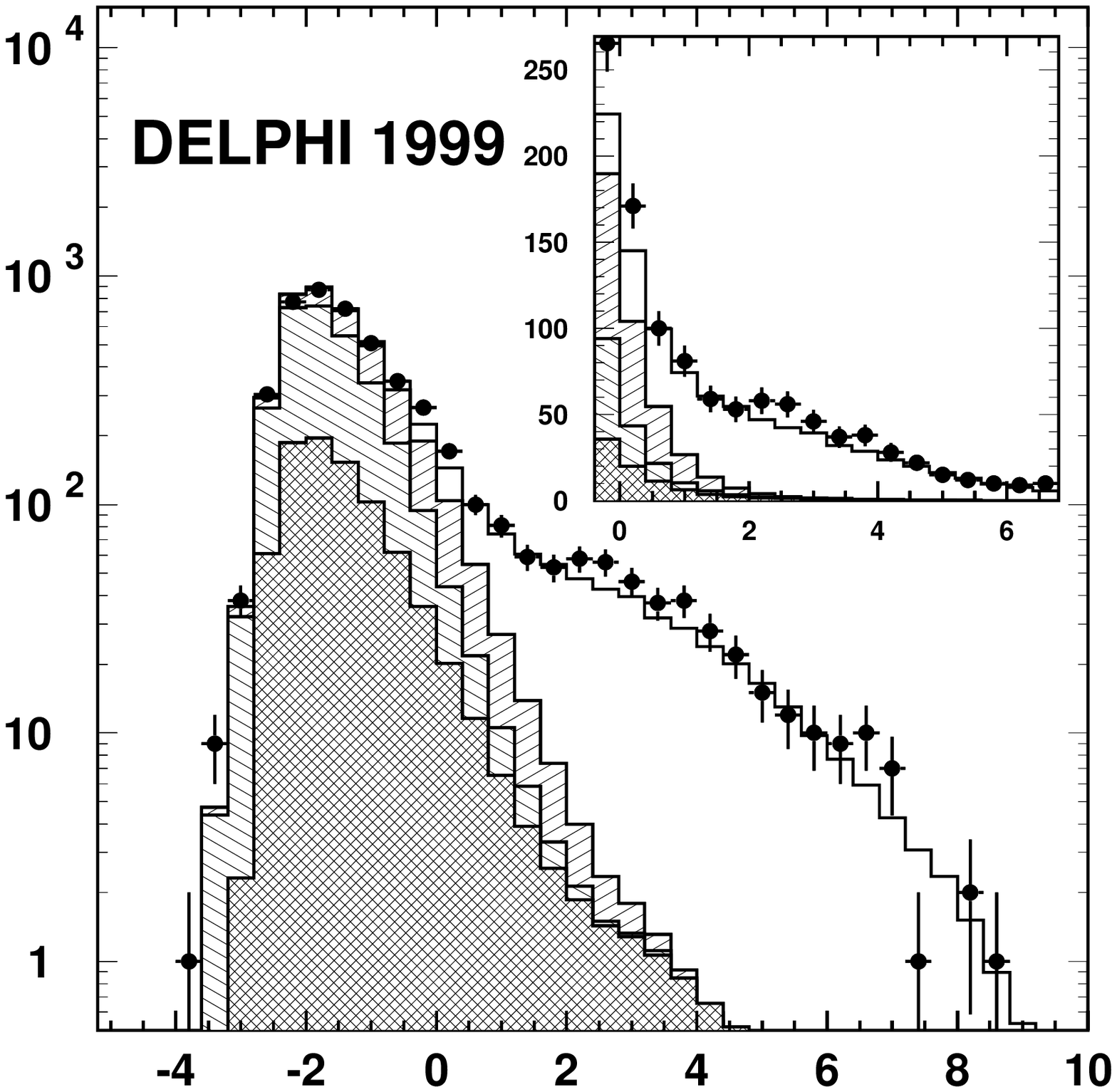,width=0.48\textwidth}
\epsfig{file=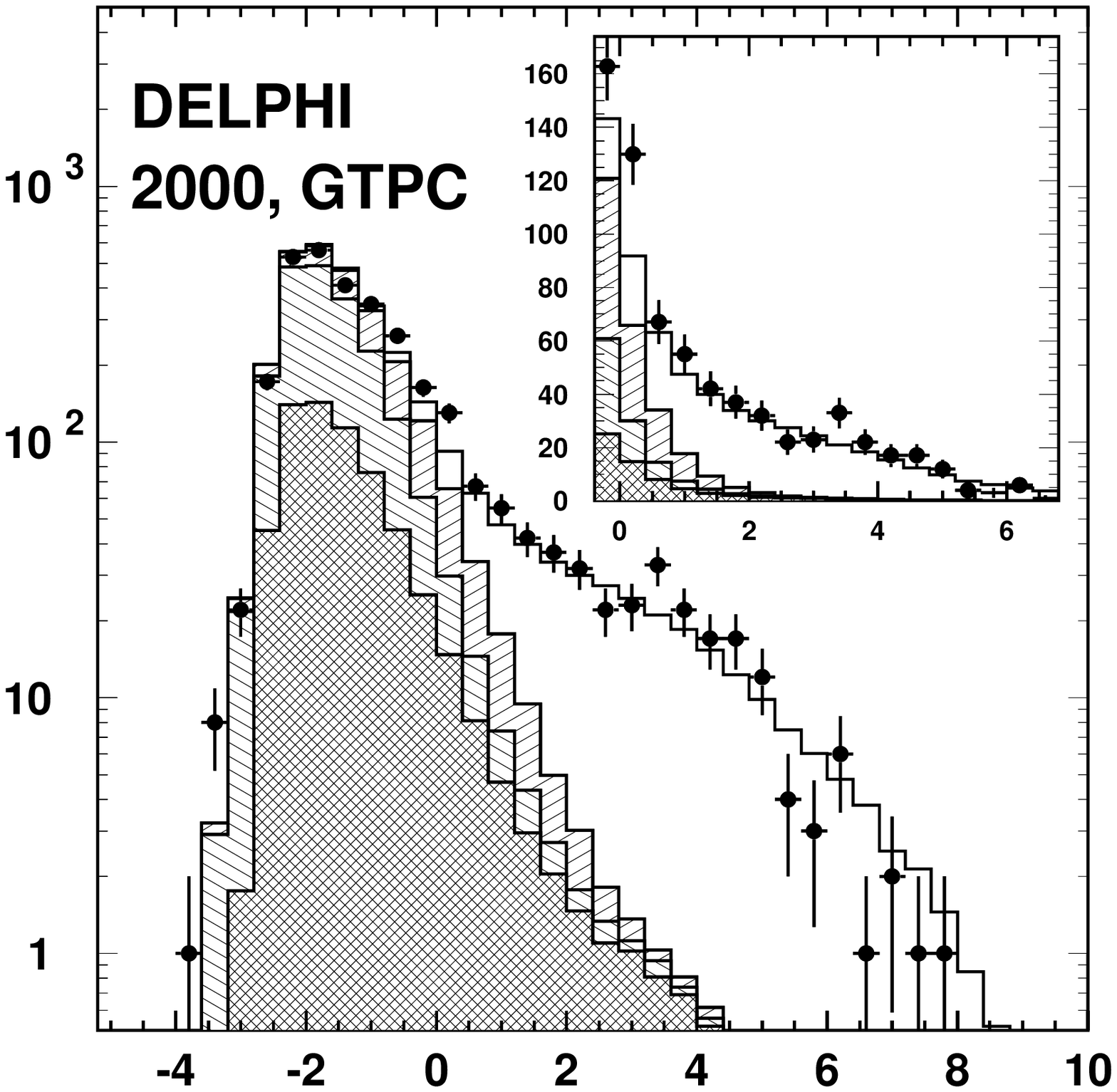,width=0.48\textwidth}
\epsfig{file=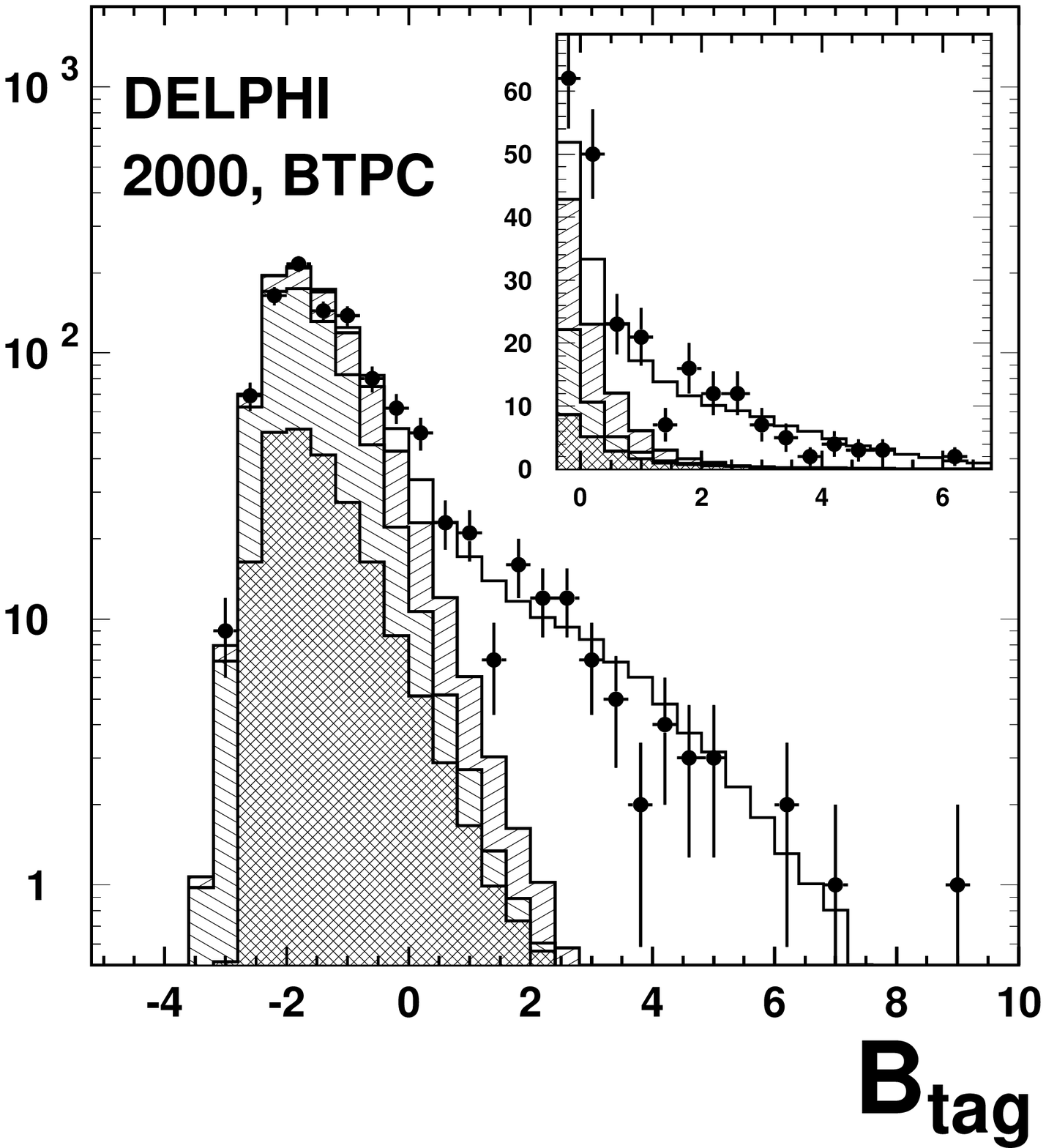,width=0.48\textwidth}
\caption[]{The variable \Bt\ plotted for all datasets.
%The points are the data,  the histogram the Monte Carlo.  No shading indicates
%\bbbar, upward diagonals represent \ccbar,  downward diagonals represent uds
%and hatching represents four-fermion (and residual $\tau^+\tau^-$) contamination.  
The standard analysis 
has a cut at \Bt\ = 1. The insets
show a zoom of the b-enhanced region on a linear scale.}
\label{fig:btag}
\end{center}
\end{figure}

The efficiency and expected background were determined primarily from Monte Carlo,
and cross-checked, where possible, from the data themselves.  
Figure~\ref{fig:btag} shows the distribution of the b-tag variable, \Bt, in data
and simulation for each dataset. In these plots the 2000 data have been
divided between GTPC and BTPC operation, and the 1995 and 1996 data have been
combined.  In general, reasonable agreement can be 
seen for all years in the region around and above the cut position of $\Bt=1.0$,
with worse agreement for the background-dominated region below the cut.
(The implications of this imperfect background description are assessed below.)

\begin{figure}
\begin{center}
\epsfig{file=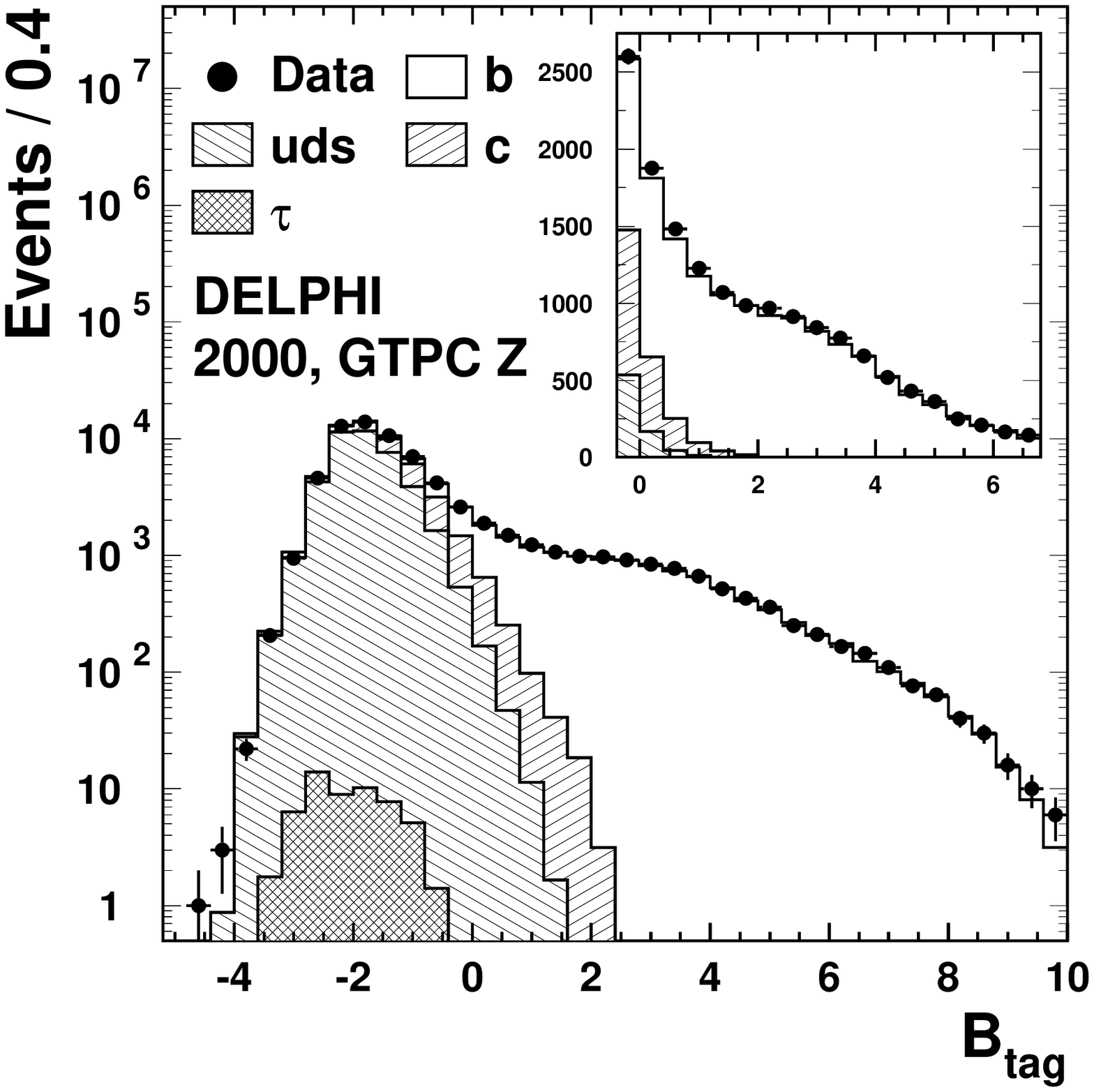,width=0.49\textwidth}
\caption[]{The variable \Bt\  for the 2000 GTPC Z-data.
The inset shows a zoom of the b-enhanced region on a linear scale.}
\label{fig:btagz}
\end{center}
\end{figure}

The running at the Z-pole in each year provides a control sample which may be used to 
calibrate the simulation. The value of \Rb\ at the Z-pole is well known 
from LEP~1~\cite{LEP1SLD}.  This value has been compared with the results obtained
from applying expression~(\ref{eq:rbextract}) to each sample of Z-calibration data.
Figure~\ref{fig:btagz} shows the distribution of \Bt\ for Z-calibration data of 
the 2000 GTPC period, 
together with that of the corresponding simulation.
The b-tag variable has a mild dependence on the collision energy.  In order
to make the Z-data study as relevant as possible to the high energy measurements,
the cut value was placed at $\Bt=0.6$ for these data, which gives a similar efficiency to the
value used at high energy.   The analysis returned a value of \Rb\ which
was similar for all datasets apart from 1998, with a mean that was $(4.1 \pm 1.2)\%$ higher 
in relative terms than the world average result.  The value found for 1998 was 
$(4.2 \pm 1.4)\%$ lower than the world average.

The offset in the measurement of \Rb\ with the Z-data can be caused by imperfections
in simulating the response of the detector to the $b$ events, the
background or to both.  (Effects arising from uncertainties in the knowledge
of the B and D decay modelling have been accounted for and found to be small.)
In order to distinguish between these possibilities, a fit was performed
to the \Bt\ distribution of the Z-data 
in the background enriched region around the cut value
($0 < \Bt <  2.5$),
taking the shapes of the signal and background from the simulation and fitting
their relative contributions.  The results returned  background scaling factors
with respect to the simulation which varied between around 0.9 and 1.2, depending on
the year, with a relative precision of better than $5\%$.  
After allowing for these corrections,  the remaining, and most
significant, cause for the offset was attributed to an incorrect estimate
of the b-tagging efficiency.

A fit was performed to the background level in the high energy data, identical
to that made with the Z-running samples.  Compatible results were obtained within
$\pm 10\%$.  For the high energy \Rb\ extraction, therefore, these Z-pole determined
scaling factors
were applied to the $\rm c\bar{c}$ and uds background, with this $10\%$ uncertainty 
assigned as a systematic error,
uncorrelated between years.
The same factors were applied to the four-fermion background,  but with twice the
systematic uncertainty, as this background component is not present in the Z-data.
Finally, the b-tagging efficiency was corrected by the amount indicated from the
low energy study, with half of this correction taken as an uncertainty, to account
for any variation with energy.  The correction factor varied between 0.959 in 1998
and 1.045 for the highest energy point of 2000.
Given the very similar nature of the offset
seen in the Z-pole study for all years apart from 1998, the uncertainty was taken as correlated
for these datasets.

The calibration procedure was repeated under different conditions and assumptions, 
for example using the same \Bt\ cut value for Z-pole and high energy data, and using an
absolute offset rather than a factor to correct the efficiency.  In all cases
compatible results were obtained.

Table~\ref{tab:makeup} shows the post b-tag sample composition at each energy point,
after applying the various corrections factors and assuming the Standard Model
production fractions.

\begin{table}
\begin{center}
\caption[]{The percentages of each event category making up the sample after the cuts on
reconstructed $\sqrt{s'/s}$ and $B_{\rm tag}$, for each energy. In the case of
\bbbar\ events the division between high and low true $\sqrt{s'/s}$ is indicated.
For the non-\bbbar\ final
states the low $\sqrt{s'}$ component is included in the category definition.  
(Note that for the energies $\sqrt{s}=$130.3-172.1~GeV, the \ccbar\ and uds background
contributions have uncertainties of around 0.5\%, due to finite Monte Carlo statistics.)}
\label{tab:makeup}
\vspace*{0.2cm}
\begin{tabular}{c|ccccc}
$\sqrt{s}$ [GeV] & $\bbbar_{\sqrt{s'/s}\ge0.85}$ &  $\bbbar_{\sqrt{s'/s}<0.85}$ & \hspace*{0.7cm}\ccbar\hspace*{0.8cm} & \hspace*{0.6cm}uds\hspace*{0.8cm} & 4-fermion \\ \hline
130.3            &       79.7           &       15.8           &  3.7   & 0.7 &     0     \\
136.3            &       77.8           &       17.9           &  2.9   & 1.4 &     0     \\
161.3            &       83.9           &       10.6           &  4.4   & 0.5 &   0.6     \\
172.1            &       82.3           &        8.4           &  4.9   & 1.6 &   2.7     \\
182.7            &       82.1           &        7.7           &  5.1   & 1.5 &   3.6     \\
188.6            &       81.8           &        7.0           &  5.6   & 1.5 &   4.2     \\
191.6            &       83.1           &        6.9           &  4.9   & 1.2 &   4.0     \\
195.5            &       82.7           &        6.7           &  4.9   & 1.5 &   4.2     \\
199.5            &       82.9           &        6.5           &  4.9   & 1.3 &   4.4     \\
201.7            &       82.6           &        6.4           &  4.8   & 1.5 &   4.6     \\
204.8            &       81.7           &        6.2           &  5.3   & 1.5 &   5.2     \\
206.6            &       82.1           &        6.0           &  5.0   & 1.6 &   5.3     \\
\end{tabular}
\end{center}
\end{table}

\subsection{Systematic Uncertainties in Modelling of Physics \\ Processes}

The stability of the results was studied with respect to uncertainties 
in the knowledge of important  properties of B and D production and decay,
and other event characteristics relevant to the b-tag.   The variation in
the parameter values was implemented by re-weighting Monte Carlo events to the modified
distribution.
\begin{itemize}
\item{{\bf b and c fragmentation:} 
Simulated \bbbar\ and \ccbar\ events at high energy had their
Peterson fragmentation parameters~\cite{PETER}
varied in the range corresponding to the uncertainties in the 
mean scaled energy of weakly decaying b and c hadrons in Z decays~\cite{LEP1SLD}.}
\item{{\bf b and c decay multiplicity: } The charged b decay multiplicity was allowed
to vary in the range  $4.955\pm0.062$~\cite{LEP1SLD} and that of D mesons
was varied according to~\cite{LEP1SLD,MARK3}, 
with a $\pm 0.5$ uncertainty assigned to the charged multiplicity
of c baryon decays. }
\item{{\bf b and c hadron composition:} The proportions of weakly decaying b and c
hadrons were varied according to the results reported in~\cite{PDG} 
and~\cite{ALEPHCHARM} respectively.}
\item{{\bf b and c hadron lifetime:} The b and c hadron lifetimes were varied within their measured range~\cite{PDG}.
In the b hadron case this was $1.576 \pm 0.016$~ps.}
%\item{{\boldmath \bf $\rm g \to \bbbar$:} 
\item{{\bf gluon splitting to heavy quarks:} 
The rate of gluon splitting to \bbbar\ and \ccbar\ per hadronic event
was varied in the range $(0.254 \pm 0.051)\%$ and $(2.96 \pm 0.38)\%$ respectively~\cite{LEP1SLD}.}
%%%%%%%%%%%%%%%%%%%%%%%%%%%%%%%%%%%%%%%%%%%%%%%%%%%%%%%%%%
\item{{\boldmath $\rm K^0_S$ \bf and $\rm \Lambda$ \bf production: } The rate of 
$\rm K^0_S$ and $\rm \Lambda$ hadrons was varied by $\pm 5 \%$, consistent 
with~\cite{DELPHIV01,DELPHIV02}. }
\end{itemize}
For each property in turn, the value of \Rb\ was recalculated using the re-weighted simulation
as input and the observed change taken as the systematic uncertainty.  
The results for the 188.6~GeV and 206.6~GeV energy points are shown in 
Table~\ref{tab:modsyst}, with the total uncertainty corresponding
to the sum in quadrature of the individual components.  
Similar behaviour was observed for the other energy points.

\begin{table}
\begin{center}
\caption[]{Fractional systematic uncertainties on \Rb\ associated with physics modelling 
for two illustrative energy points. Values are given in percent.}
\label{tab:modsyst}
\vspace*{0.2cm}
\begin{tabular}{l|cc} 
                     & \multicolumn{2}{c}{Energy point} \\
Uncertainty Source   &   188.6 GeV            &    206.6 GeV     \\ \hline
b fragmentation      & \hspace{0.3cm} $0.2$  & \hspace{0.3cm} $0.2$ \\
b decay multiplicity & \hspace{0.3cm} $0.5$  & \hspace{0.3cm} $0.7$ \\
b hadron composition & \hspace{0.3cm} $0.2$  & \hspace{0.3cm} $0.2$ \\
b hadron lifetime    & \hspace{0.3cm} $0.2$  & \hspace{0.3cm} $0.3$ \\
c fragmentation      & \hspace{0.3cm} $0.1$  & \hspace{0.3cm} $0.1$ \\
c decay multiplicity & \hspace{0.3cm} $0.3$  & \hspace{0.3cm} $0.2$ \\
c hadron composition & \hspace{0.3cm} $0.2$  & \hspace{0.3cm} $0.2$ \\
c lifetime           & \hspace{0.3cm} $0.1$  & \hspace{0.3cm} $0.1$ \\
$\rm g \to \bbbar$   & \hspace{0.3cm} $0.1$  & \hspace{0.3cm} $0.1$ \\
$\rm g \to \ccbar$   & \hspace{0.3cm} $0.1$                & $<0.1$ \\
$\rm K^0_S$ and $\Lambda$ production 
                     & \hspace{0.3cm} $0.2$  & \hspace{0.3cm} $0.3$ \\ \hline
Total                & \hspace{0.3cm} $0.8$  & \hspace{0.3cm} $0.9$ \\ 
\end{tabular}
\end{center}
\end{table}

\subsection{Summary of Systematics and Results}
\label{sec:rbresults}

The relative systematic uncertainties on \Rb\ are summarised in 
Table~\ref{tab:systsumrb}.   
In addition to those components already discussed,
contributions are included which arise from the finite size of the 
Monte Carlo simulation sample, and from the effect of the uncertainty
in the residual radiative contamination in the analysis.
Studies on the resolution of the  $\sqrt{s'/s}$
reconstruction indicated that this background was understood to the level of $10\%$.
It can be seen that the dominant source of systematic uncertainty 
is that coming from the comparison with the Z-data. 
  
The results for \Rb\ are given in Table~\ref{tab:rbres}, together
with the statistical and systematic uncertainties.
The correlation matrix for these results can be found in Appendix~\ref{sec:cormat}.
For each of the two energy points of the year 2000 the results for the GTPC and BTPC period
are found to be compatible and are thus combined into a single value.
No variation
of $R_{\rm c}$ is considered in the systematic uncertainty, but 
the dependence of $R_{\rm b}$ on this quantity, $\Delta R_{\rm b}/(R_{\rm c} \,-\, R^{\rm SM}_{\rm c})$, 
is tabulated explicitly.

\begin{table}
\begin{center}
\caption[]{The fractional systematic uncertainty, in percent, on \Rb, energy point by energy point.}
\label{tab:systsumrb}
\vspace*{0.2cm}
\begin{tabular}{c|ccccc|c}
$\sqrt{s}$ [GeV] &  Z Comparison & Modelling &  4-fermion & MC Stats &  Rad. Bckgd. & \hspace*{0.2cm}  Total \hspace*{0.2cm}  \\ \hline
130.3            &      1.7      &    1.1    &     /      &   2.4    &     0.5         & 3.2   \\
136.3            &      1.8      &    1.1    &     /      &   2.9    &     0.4         & 3.6   \\
161.3            &      1.6      &    1.1    &    0.1     &   1.9    &     0.1         & 2.7   \\
172.1            &      1.8      &    1.1    &    0.5     &   2.2    &     0.1         & 3.1   \\
182.7            &      2.1      &    1.0    &    0.9     &   0.4    &     0.1         & 2.5   \\
188.6            &      2.1      &    0.8    &    0.8     &   0.4    &     0.1         & 2.4   \\
191.6            &      2.2      &    0.8    &    0.8     &   0.5    &     0.3         & 2.5   \\
195.5            &      2.2      &    0.9    &    0.9     &   0.5    &     0.1         & 2.6   \\
199.5            &      2.2      &    1.0    &    0.9     &   0.5    &     0.2         & 2.6   \\
201.7            &      2.2      &    0.9    &    0.9     &   0.4    &     0.2         & 2.6   \\
204.8            &      2.1      &    0.9    &    1.2     &   0.4    &     0.1         & 2.6   \\
206.6            &      2.4      &    0.9    &    1.1     &   0.3    &     0.2         & 2.8   \\
\end{tabular}
\end{center}
\end{table}

\begin{table}
\begin{center}
\caption[]{The results for \Rb\ at each energy point.  Also given are the dependences of  \Rb\ 
on $R_{\rm c}$, and the values for the latter fraction assumed in the analysis~\cite{ZFIT}. For convenience, the
corresponding Standard Model expectations for \Rb\ are included.}
\label{tab:rbres}
\vspace*{0.2cm}
\begin{tabular}{c|ccccc|ccl}
%$\sqrt{s}$ [GeV] & \Rb & & stat & & syst & $\frac{\Delta R_{\rm b}}{(R_{\rm c} \,-\, R^{\rm SM}_{\rm c})}$ & $R^{\rm SM}_{\rm c}$ & $R^{\rm SM}_{\rm b}$ 
$\sqrt{s}$ [GeV] & \Rb & & $\sigma_{\rm stat}$ & & $\sigma_{\rm syst}$ & $\frac{\Delta R_{\rm b}}{(R_{\rm c} \,-\, R^{\rm SM}_{\rm c})}$ & $R^{\rm SM}_{\rm c}$ & $R^{\rm SM}_{\rm b}$ 
 \\ \hline
130.3 & 0.228 &$\pm$& 0.041 &$\pm$&0.007 &  $-$0.027 & 0.220 & 0.186 \\
136.3 & 0.153 &$\pm$& 0.041 &$\pm$&0.006 &  $-$0.023 & 0.226 & 0.182 \\
161.3 & 0.183 &$\pm$& 0.029 &$\pm$&0.005 &  $-$0.023 & 0.244 & 0.170 \\
172.1 & 0.127 &$\pm$& 0.028 &$\pm$&0.004 &  $-$0.023 & 0.249 & 0.167 \\
182.7 & 0.127 &$\pm$& 0.013 &$\pm$&0.003 &  $-$0.032 & 0.253 & 0.165 \\
188.6 & 0.166 &$\pm$& 0.009 &$\pm$&0.004 &  $-$0.035 & 0.255 & 0.164 \\
191.6 & 0.194 &$\pm$& 0.024 &$\pm$&0.005 &  $-$0.032 & 0.256 & 0.163 \\
195.5 & 0.161 &$\pm$& 0.013 &$\pm$&0.004 &  $-$0.031 & 0.258 & 0.163 \\
199.5 & 0.187 &$\pm$& 0.014 &$\pm$&0.005 &  $-$0.031 & 0.258 & 0.162 \\
201.7 & 0.183 &$\pm$& 0.020 &$\pm$&0.005 &  $-$0.030 & 0.259 & 0.162 \\
204.8 & 0.156 &$\pm$& 0.014 &$\pm$&0.004 &  $-$0.031 & 0.259 & 0.161 \\
206.6 & 0.163 &$\pm$& 0.011 &$\pm$&0.005 &  $-$0.029 & 0.260 & 0.161 \\
\end{tabular}
\end{center}
\end{table}

The internal consistency of the measured \Rb\ results may be studied,
under the assumption that any dependence of the true value
on collision energy can be neglected.  The pull distribution of $(\Rb - <\Rb>)/\sigma$ 
is found to have a spread of 1.2, with the most outlying 
entry arising from the measurement at $\sqrt{s} =$~183~GeV,
which is 2.7 $\sigma$ below the mean.

The stability of the results has been examined when
changing the value of the b-tag cut.  The cut position was tightened to a value
of \Bt=2.5 in the high energy data, and \Bt=2.1 in the Z-data,
and \Rb\ re-evaluated at each energy point.
Under this selection the event samples halve in
size, but the non-\bbbar\ background is reduced by
almost a factor of three.   No statistically significant
change in result was observed with respect to the standard
selection for any energy point in isolation, nor 
for all energy points averaged together, indicating that
the background levels and efficiency are well understood 
for both selections.

The results for \Rb\ are compared with the Standard Model expectations and
interpreted in the context of possible new physics contributions in
Sect.~\ref{sec:interpret}.

\section{Measurement of \Ab}
\label{sec:afb}

\subsection{Procedure}
\label{sec:afbproc}

For the non-radiative \bbbar\ events selected in this study,  the expected
form of the differential cross-section is given by:
\begin{eqnarray}
\frac{d \sigma_{\rm b}} {d \cos \theta_{\rm b}} 
& \propto & 
1 \, + \, \cos^2\theta_{\rm b} \, + \, \frac{8}{3} \Ab\cos \theta_{\rm b},
\label{eq:diffxsec}
\end{eqnarray}
where $\theta_{\rm b}$ is the polar angle the $\rm b$-quark 
makes with the initial $e^-$ direction. 

\begin{figure}
\begin{center}
\epsfig{file=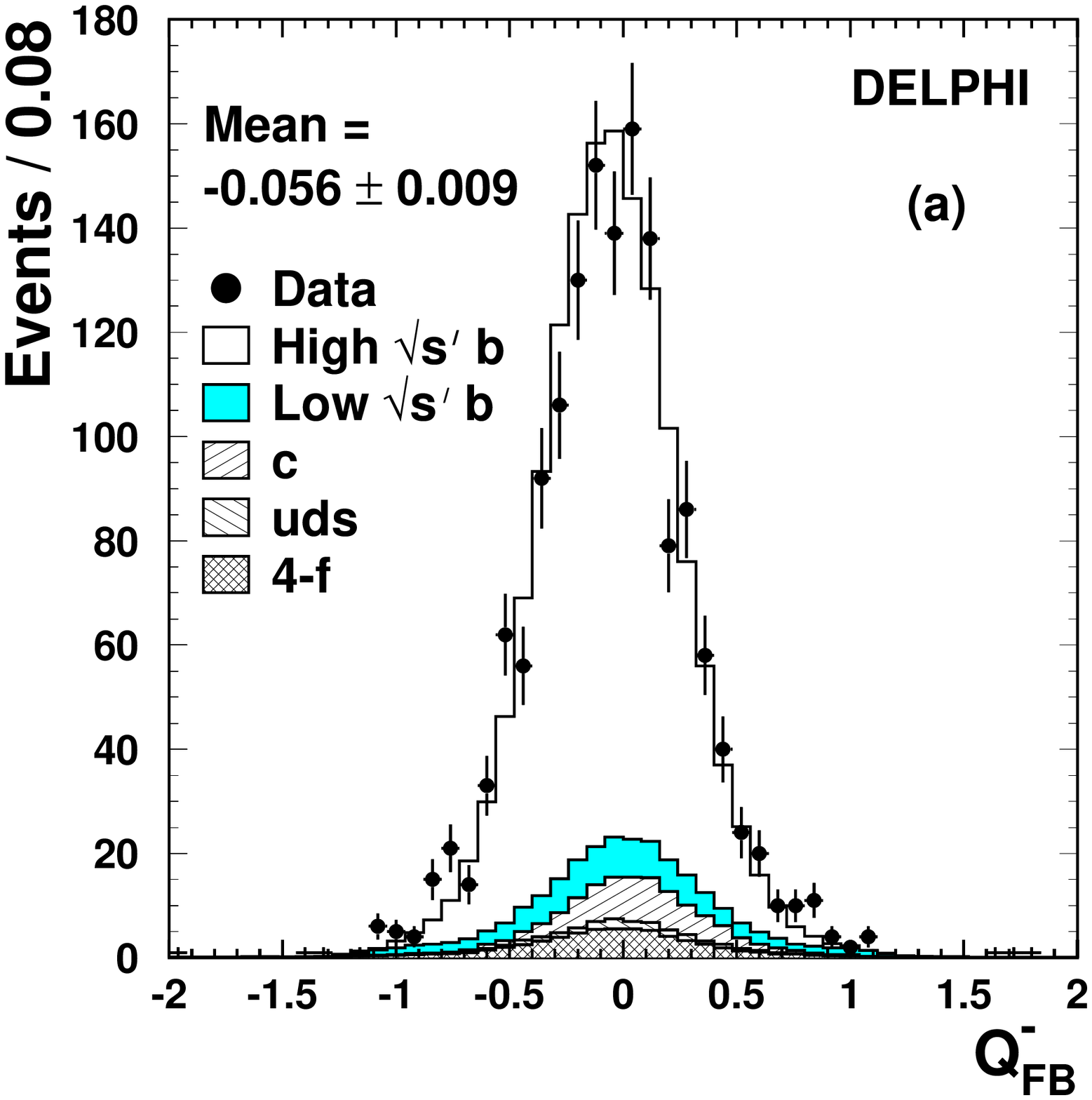,width=0.45\textwidth}
\epsfig{file=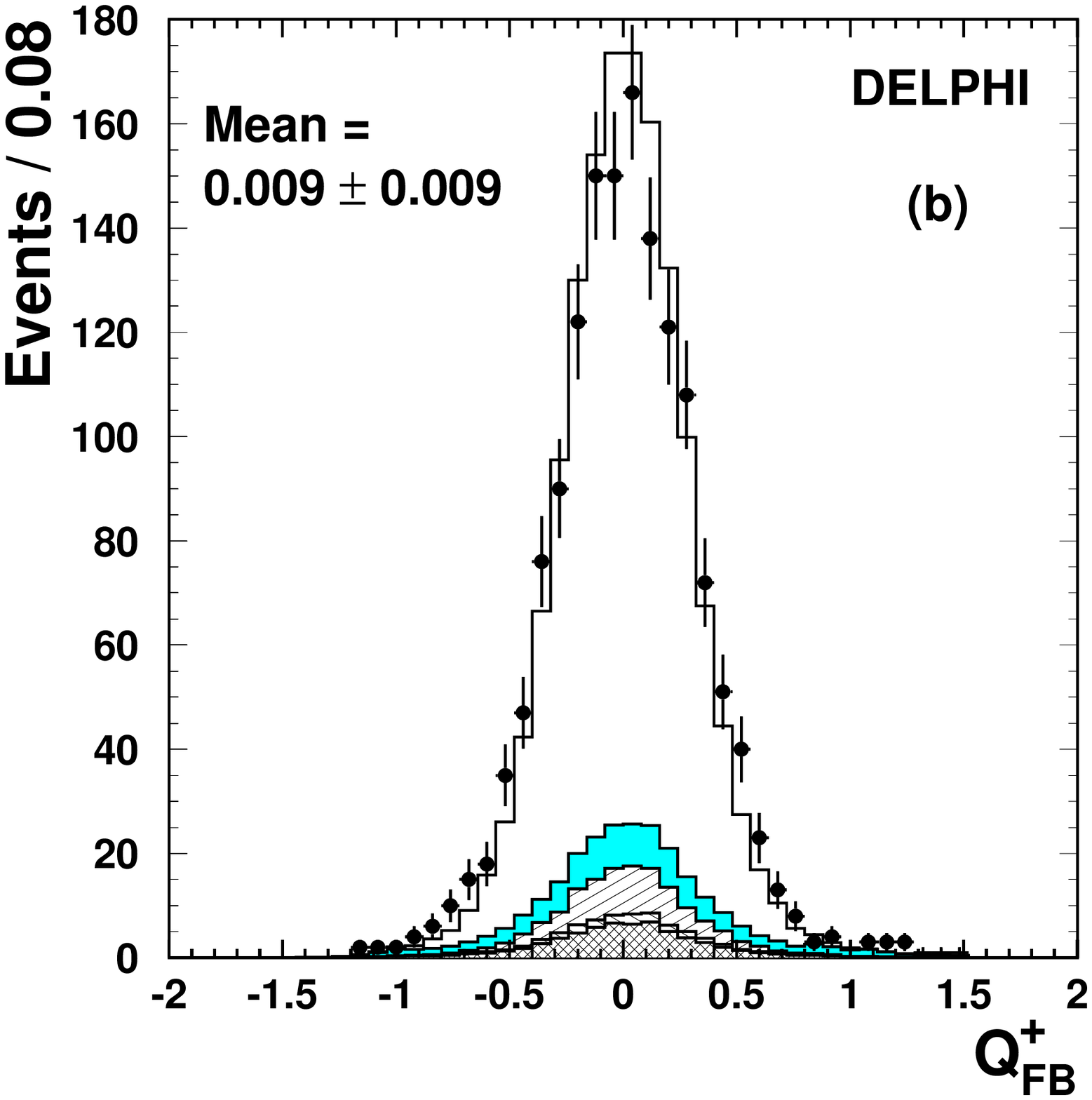,width=0.45\textwidth}
\caption{Distribution of the two event charge variables for all data after b-tag cut.  
(a) shows the charge asymmetry between the two hemispheres, $Q^-_{\rm FB}$.  (b) shows the sum of 
the hemisphere charges, $Q^+_{\rm FB}$. Also shown are the expectations from the simulation, 
which are generated with the Standard Model values for the asymmetries of each
component.}
\label{fig:qfqb}
\end{center}
\end{figure}

The analysis presented in this paper is based on an unbinned likelihood
fit to expression~(\ref{eq:diffxsec}), and hence requires knowledge of
$\theta^{\rm rec}_{\rm b}$, which is the event-by-event value 
of $\theta_{\rm b}$ as reconstructed in DELPHI.   This
reconstruction is performed using the thrust axis and a hemisphere
charge technique.   
Each event is divided into two hemispheres by the plane perpendicular to the thrust axis
that contains the nominal interaction point.   Simulation shows that for non-radiative events
the thrust axis is a good approximation to the direction of emission of the initial \bbbar\ pair.
Then the `hemisphere charges' $Q_{\rm F}$ and $Q_{\rm B}$ are calculated for the forward and backward
hemispheres. $Q_{\rm F}$ is defined:
\begin{eqnarray}
Q_{\rm F   } &\equiv& \frac{ \sum_i q_i | {\bf p}_i \cdot {\bf T} |^\kappa}
                  { \sum_i       | {\bf p}_i \cdot {\bf T} |^\kappa},
\label{eq:qf}
\end{eqnarray}
where ${\bf p}_i$ and $q_i$ are the momentum and charge of particle $i$, $\bf T$ is the thrust axis,
$\kappa$ is an empirical parameter, and the sum runs over all charged particle tracks
for which ${\bf p}_i \cdot {\bf T} > 0$. $Q_{\rm B}$ is defined in an analogous manner with the 
requirement that ${\bf p}_i \cdot {\bf T} < 0$.   The information from both hemispheres
may be combined into two event variables:
\begin{eqnarray}
Q^\pm_{\rm FB} & \equiv & Q_{\rm F} \pm Q_{\rm B}.
\end{eqnarray}
The sign of $Q^-_{\rm FB}$ is sensitive to whether the b-quark was emitted in the forward or backward
hemisphere.  The value of $\kappa$ in equation~(\ref{eq:qf}) is tuned to maximise this
discrimination, and is set to 0.5.   Figure~\ref{fig:qfqb}~(a) shows $Q^-_{\rm FB}$, plotted for
all data.  There is a small, but significant negative offset, indicating that the b-quark
is preferentially emitted in the forward hemisphere.  $Q^+_{\rm FB}$ has no sensitivity to
the initial b-quark direction, but provides a quantity which can be compared between data
and simulation, with a width that reflects the resolution of the method.
$Q^+_{\rm FB}$ is
plotted in  Fig.~\ref{fig:qfqb}~(b), together with the corresponding quantity from the 
simulation.  As expected, it is centred on zero.  
The distribution is marginally wider
in data than in the Monte Carlo.

\begin{figure}[htb]
\begin{center}
\epsfig{file=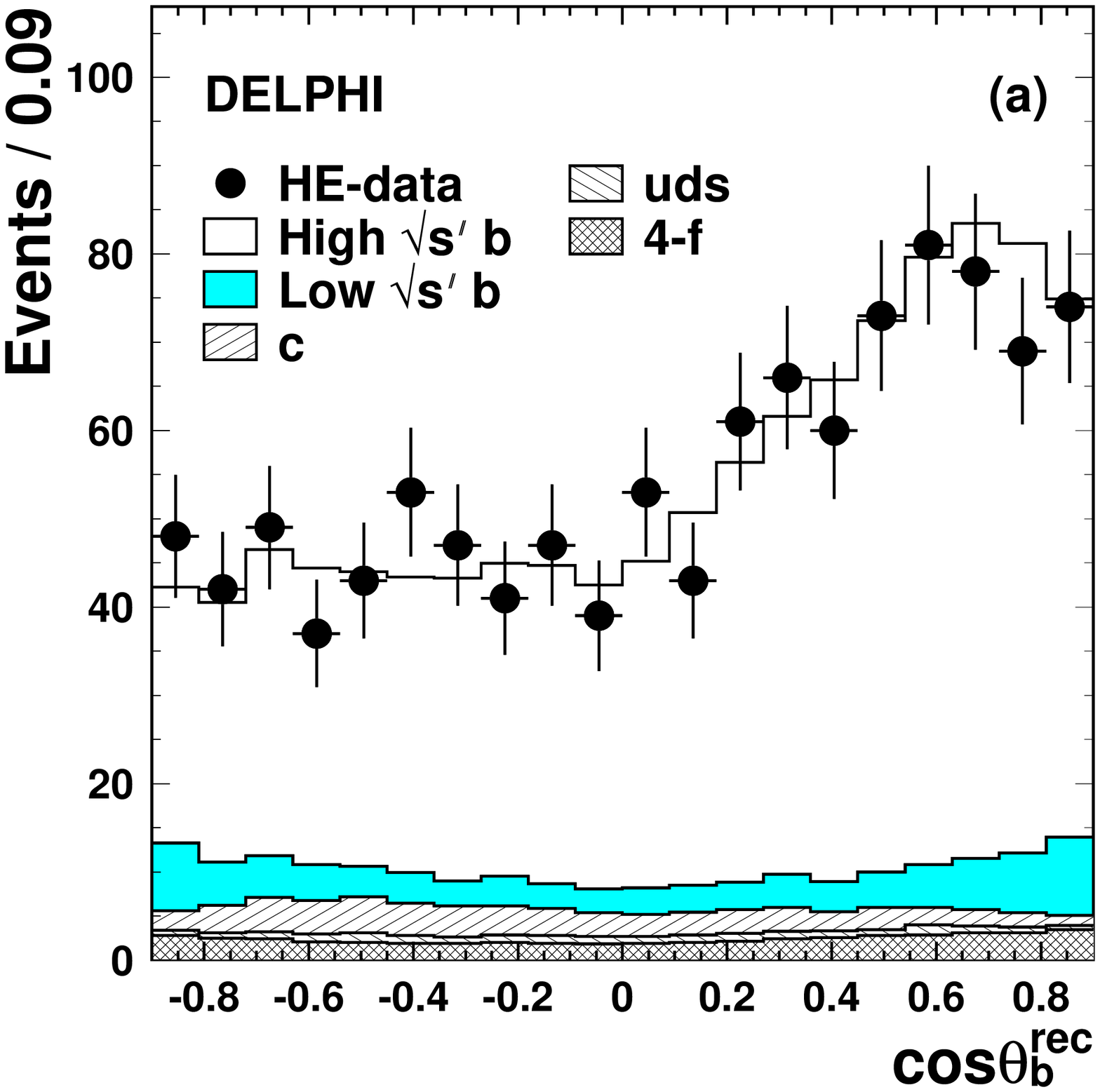,width=0.45\textwidth}
\epsfig{file=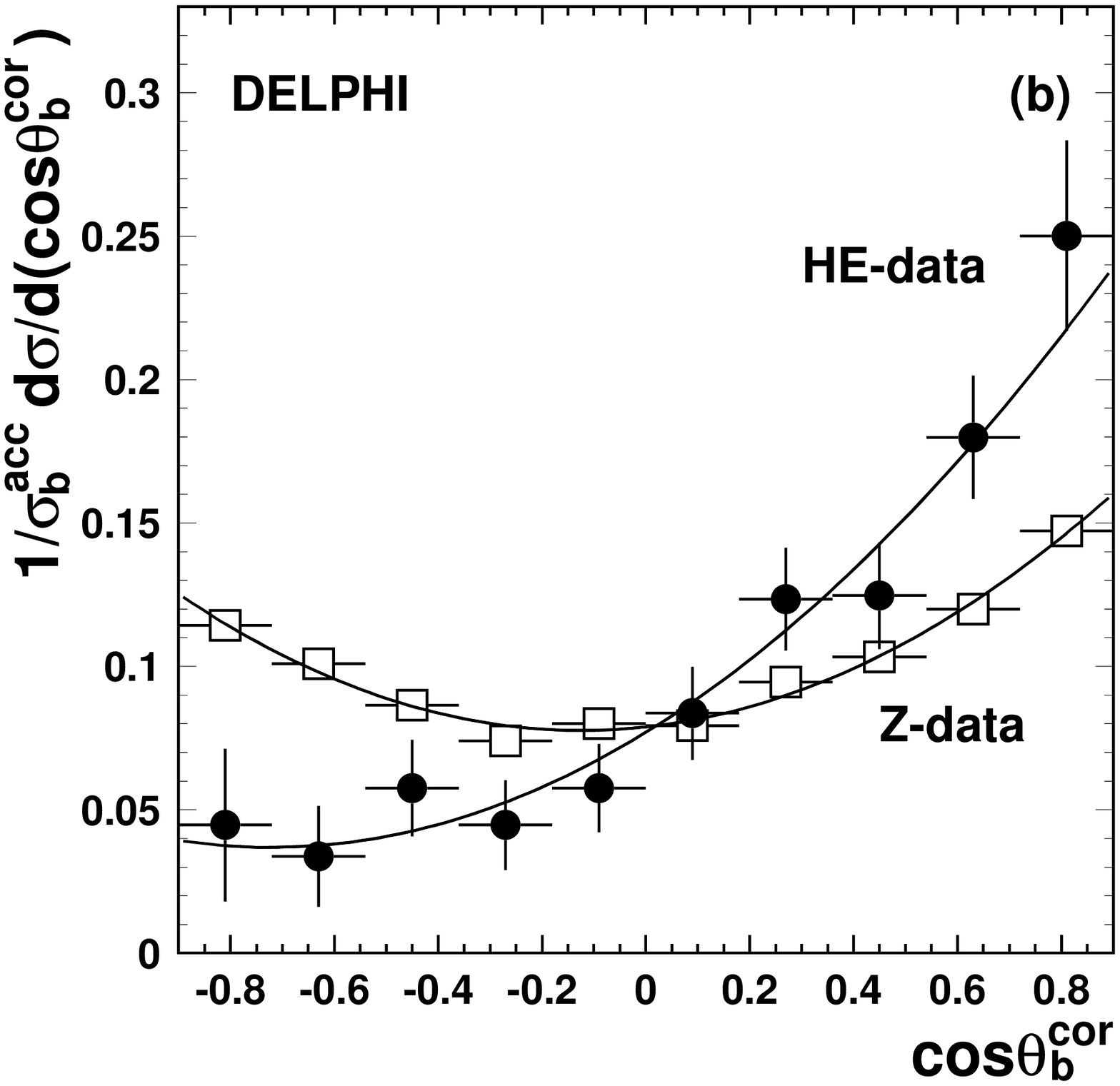,width=0.45\textwidth}
\caption{The observed angular distribution for all data after b-tag cut and the 
requirement  $|Q^-_{\rm FB}|>0.1$.  (a) shows the raw distribution of
events with respect to $\rm \cos \theta_b^{rec}$ together with the
expectations from simulation,  generated with the Standard Model 
values for the asymmetries of each component.  (b) shows the differential
cross-section (normalised to the total cross-section within the acceptance) with respect
to $\rm \cos \theta_b^{cor}$, where $\rm \theta_b^{cor}$ is the b-quark direction 
after correction for wrong flavour tags, non-uniform acceptance efficiency and background.
Also shown is the corresponding distribution for the LEP~2 Z-data.
The superimposed curves are fits to the form of the expected differential cross-section.}
\label{fig:cosrec}
\end{center}
\end{figure}

The cosine of the reconstructed b-quark direction is then given by:
\begin{equation}
X \equiv \cos \theta^{\rm rec}_{\rm b} =  {- \: \rm sign}(Q^-_{\rm FB}) \cdot |\cos \theta_T |,
\end{equation}
where $\theta_T$ is the polar angle of the thrust axis.  The distribution
of  $\cos \theta^{\rm rec}_{\rm b}$ is shown in Fig.~\ref{fig:cosrec}~(a),
for the full LEP~2 dataset, plotted for events where $|Q^-_{\rm FB}|> 0.1$.
%For comparison, the same variable is plotted as reconstructed on the Z-data.   
The asymmetry which is observed  
is an underestimate of the real asymmetry, both
because of `mistags' and because of background contamination.  Detector inefficiencies
also distort the distributions, particularly in the forward and backward regions.
Mistags are events in which the sign of $Q^-_{\rm FB}$ does not give the correct 
b-quark direction. Mistags dilute the true asymmetry by a factor $D = (1-2\omega)$, 
where $\omega$ is the
probability of mistag. Note that $\omega$ has a dependence on the absolute value 
of $Q^-_{\rm FB}$.  For example, simulation indicates that for the ensemble of high
energy data the mistag rate has a value of $\omega = 0.45$ for events where $|Q^-_{\rm FB}|< 0.1$,
and $\omega = 0.27$ in the case when $|Q^-_{\rm FB}|> 0.1$, falling to
$\omega = 0.17$ when  $|Q^-_{\rm FB}|> 0.36$.   
Figure~\ref{fig:cosrec}~(b) shows
the same data after correction for background contamination, detector inefficiency and
mistags,
and the corresponding distribution for the Z-data. 
It is apparent that the high energy data exhibit an asymmetry significantly 
higher than that of the Z-data, which have a value consistent with that measured at LEP~1~\cite{LEP1SLD}.

Optimal sensitivity to \Ab\ is achieved through performing a
maxmimum likelihood fit, taking as the probability density function the expected 
differential cross-section of equation~(\ref{eq:diffxsec}).   At each energy point, the
measured asymmetry $A^{\rm meas}_{\rm FB}$ is determined by maximising the following expression:
\begin{equation}
\ln {\cal L} = \sum_i \ln \left( 1 \, + \, (X_i)^2 \, + \, \frac{8}{3} A^{\rm meas}_{\rm FB} X_i \right),
\label{eq:like}
\end{equation}
where the sum runs over all events.
Mistags and contamination are accounted for by writing 
\begin{equation}
A^{\rm meas}_{\rm FB} = \sum_j f_j D_j A_j.
\end{equation}
Here the sum runs over the five categories of event type in the 
sample: signal, radiative
\bbbar\ contamination, \ccbar, light quark and four-fermion.  Each category enters with a
proportion $f_j$, as given by the values in Table~\ref{tab:makeup}, with a true asymmetry $A_j$
and dilution factor $D_j$, where $A_j$ for the signal category is equivalent to \Ab.
For the purposes of accounting for the background in the fit, equation~(\ref{eq:diffxsec}) is 
an adequate description of the distribution of radiative and four-fermion events.
The dilution factors are determined from simulation, and the asymmetries of the background
processes are set to their Standard Model expectations.  
In order to exploit the dependence of the mistag probability on the absolute value
of the charge asymmetry, all events are used, 
but the dilutions and event fractions are evaluated in four bins 
of $|Q^-_{\rm FB}|$ and included in the fit accordingly.

The fit procedure has been tested on a large ensemble of simulated experiments, and found
to give unbiased results with correctly estimated uncertainties.  It has also been applied to
the Z-data. Averaged over all datasets, the measured asymmetry minus that value 
determined at LEP~1~\cite{LEP1SLD} is found to be $-0.01 \pm 0.01$.

\subsection{Results and Systematic Uncertainties}
\label{sec:afbresults}

The most important source of systematic uncertainty in the asymmetry measurement
is associated with the knowledge of the performance of the charge asymmetry 
variable.  There are three significant contributions to this uncertainty:
\begin{itemize}
\item{{\bf Detector Response:}  The distribution of track multiplicity as a function
of momentum has small differences between data and Monte Carlo both at high and low 
momentum, which may be attributed to an imperfect modelling of the track reconstruction
in the simulation.   Tracks were re-weighted in the simulation in order to establish
the effect on the mistag rate.  Similar studies were conducted to understand the consequences
of differences in the momentum resolution between data and Monte Carlo.   Finally,
the width of the $Q^+_{\rm FB}$ distribution was artificially increased in the simulation, to match
that of the data, by adjusting the value of the $\kappa$ parameter in the analysis of 
the simulation alone, and the effect on $Q^-_{\rm FB}$ was determined.}
\item{{\bf Hadronisation:} An alternative Monte Carlo data set of events based on
ARIADNE~\cite{ARIADNE} was used to assess the robustness of the estimation of the 
mistag rate with respect to the description of the hadronisation process used
in the simulation.}
\item{{\bf Monte Carlo Statistics:} The limited amount of simulation data available 
introduces a non-negligible statistical uncertainty in the knowledge of the mistag rate.}
\end{itemize}
Additional possible sources of measurement bias related to the mistag have been considered, for example whether
any significant angular dependence exists in the value of the dilution.
These effects were found to have negligible impact on the results.

In addition to these studies, systematic uncertainties were evaluated arising from the 
same three sources that were considered in the \Rb\ measurement, namely
the uncertainty associated with the sample composition as assessed from
the Z-data; the uncertainty in the level of the
4-fermion background; and the uncertainty in the modelling of the physics processes
(apart from hadronisation).    
The modelling systematic here includes a component
arising from the uncertainty in the knowledge of 
the b-mixing parameter $\chi$.  This was varied within the range $0.128 \pm 0.008$, 
following the evaluation reported in~\cite{PDG}.  A further uncertainty is
assigned to account for the fact that QCD corrections to the final state,
in particular gluon radiation, modify the asymmetry.  The size of this
effect has been estimated using ZFITTER~\cite{ZFIT} to be 0.018.  In practice
the selection cuts disfavour events with hard gluon radiation and thus
will suppress this correction.  In this study, however, the full effect 
is taken as an uncertainty, fully correlated between energy points. 
Finally, a systematic error is 
added to account for the uncertainty in the  knowledge of the residual radiative \bbbar\ 
contamination in the sample.  
%Studies on the resolution of the  $\sqrt{s'/s}$
%reconstruction indicated that this background was understood to the level of $10\%$.

Table~\ref{tab:afbsyst} lists the systematic uncertainties for the 188.6~GeV and 
206.6~GeV energy points.   The total is the sum in quadrature of the uncorrelated
component uncertainties.
The results for \Ab, including statistical and
systematic uncertainties, are shown in Table~\ref{tab:afbres}.  
%The central value at 161.3~GeV lies outside the physically allowed region,
%although by less than one standard deviation. 
The correlation matrix for these results can be found in Appendix~\ref{sec:cormat}.
Both the statistical uncertainty
and certain components of the systematic uncertainty have a dependence on the absolute
value of the asymmetry.  The uncertainties shown have been evaluated assuming the 
Standard Model value.

\begin{table}
\begin{center}
\caption[]{Systematic uncertainties on \Ab\ for two illustrative energy points.}
\label{tab:afbsyst}
\vspace*{0.2cm}
\begin{tabular}{l|cc} 
                     & \multicolumn{2}{c}{Energy point} \\
Uncertainty Source   &  188.6~GeV    &    206.6~GeV  \\ \hline
Detector Response    &   $0.054$      &    $0.038$      \\
Hadronisation        &   $0.027$      &    $0.025$      \\
MC Statistics        &   $0.016$      &    $0.011$      \\ 
Z Comparison         &   $0.008$      &    $0.004$      \\
Modelling            &   $0.008$      &    $0.008$      \\
QCD Correction       &   $0.018$      &    $0.018$      \\ 
4-fermion            &   $0.003$      &    $0.006$      \\
Radiative background &   $0.004$      &    $0.004$      \\ \hline
Total                &   $0.066$      &    $0.051$      \\
\end{tabular}
\end{center}
\end{table}

\begin{table}
\begin{center}
\caption[]{The results for \Ab\ at each energy point, together with the Standard Model
expectation~\cite{ZFIT}.}
\label{tab:afbres}
\vspace*{0.2cm}
\begin{tabular}{c|rcccc|c  }
$\sqrt{s}$ [GeV] & \Ab & & $\sigma_{\rm stat}$ & & $\sigma_{\rm syst}$ & $A^{\rm b, \, SM}_{\rm FB}$ \\ \hline
%130.3 & 0.569 &$\pm$& 0.507 &$\pm$&0.111 &  0.473 \\
%136.3 & 0.447 &$\pm$& 0.615 &$\pm$&0.115 &  0.496 \\
%161.3 & 1.344 &$\pm$& 0.346 &$\pm$&0.096 &  0.550 \\
%172.1 & 0.407 &$\pm$& 0.523 &$\pm$&0.097 &  0.564 \\
%182.7 &-0.120 &$\pm$& 0.245 &$\pm$&0.101 &  0.575 \\
%188.6 & 0.703 &$\pm$& 0.157 &$\pm$&0.064 &  0.579 \\
%191.6 & 0.391 &$\pm$& 0.304 &$\pm$&0.046 &  0.582 \\
%195.5 & 0.875 &$\pm$& 0.221 &$\pm$&0.057 &  0.584 \\
%199.5 & 0.602 &$\pm$& 0.185 &$\pm$&0.049 &  0.587 \\
%201.7 & 0.756 &$\pm$& 0.298 &$\pm$&0.052 &  0.588 \\
%204.8 & 0.718 &$\pm$& 0.252 &$\pm$&0.058 &  0.590 \\
%206.6 & 0.108 &$\pm$& 0.180 &$\pm$&0.048 &  0.591 \\
%% Now with QCD error added.
130.3 & 0.569 &$\pm$& 0.507 &$\pm$&0.112 &  0.473 \\
136.3 & 0.447 &$\pm$& 0.615 &$\pm$&0.117 &  0.496 \\
161.3 & 1.344 &$\pm$& 0.346 &$\pm$&0.097 &  0.550 \\
172.1 & 0.407 &$\pm$& 0.523 &$\pm$&0.099 &  0.564 \\
182.7 &-0.120 &$\pm$& 0.245 &$\pm$&0.102 &  0.575 \\
188.6 & 0.703 &$\pm$& 0.157 &$\pm$&0.066 &  0.579 \\
191.6 & 0.391 &$\pm$& 0.304 &$\pm$&0.049 &  0.582 \\
195.5 & 0.875 &$\pm$& 0.221 &$\pm$&0.060 &  0.584 \\
199.5 & 0.602 &$\pm$& 0.185 &$\pm$&0.052 &  0.587 \\
201.7 & 0.756 &$\pm$& 0.298 &$\pm$&0.055 &  0.588 \\
204.8 & 0.718 &$\pm$& 0.252 &$\pm$&0.061 &  0.590 \\
206.6 & 0.108 &$\pm$& 0.180 &$\pm$&0.051 &  0.591 \\
\end{tabular}
\end{center}
\end{table}

The self-consistency of the results may be assessed assuming that
any dependence of the true value of \Ab\ on the collision energy can
be neglected.  The pull distribution of $(\Ab \,-\, < \Ab >)/\sigma$ is
found to have a spread of 1.5.    The outliers contributing to 
this larger than expected width are the dataset at 161.3~GeV,
which has an asymmetry which is 2.3 $\sigma$ higher than the mean, and the samples at 
182.7~GeV and 206.6~GeV, which have asymmetries that are low by 2.7 and 2.4 $\sigma$ 
respectively.  The 206.6~GeV dataset is made
up of events accumulated during both the GTPC and  BTPC running;  the values of the
asymmetry and associated statistical uncertainties are found to
be $0.087 \pm 0.218$ and $0.152 \pm 0.318$, and hence consistent, for the
two periods.    All asymmetries have been re-evaluated with a more 
severe b-tag cut of 2.5, as was done for the \Rb\ analysis.  
Averaged over all data points the asymmetry
is found to shift by $-0.008 \pm 0.052$ with respect to the central
values reported in Table~\ref{tab:afbres}.   The shifts for the 161.3~GeV, 182.7~GeV
and 206.6~GeV datasets are $0.019 \pm 0.209$, $-0.278 \pm 0.191$ and $-0.043 \pm 0.162$
respectively. The magnitudes and signs of these changes do not suggest that there is
any significant problem with the understanding of the background level and behaviour.
Further cross-checks were performed in which the fit was restricted to 
high values of $|Q^-_{\rm FB}|$ and where alternative methods,
such as a binned least-squared fit, were used to determine the asymmetry.
Again, no significant changes were observed in the results, in particular
those of the three outlying points.

\section{Interpretation}
\label{sec:interpret}

\begin{figure}
\begin{center}
\epsfig{file=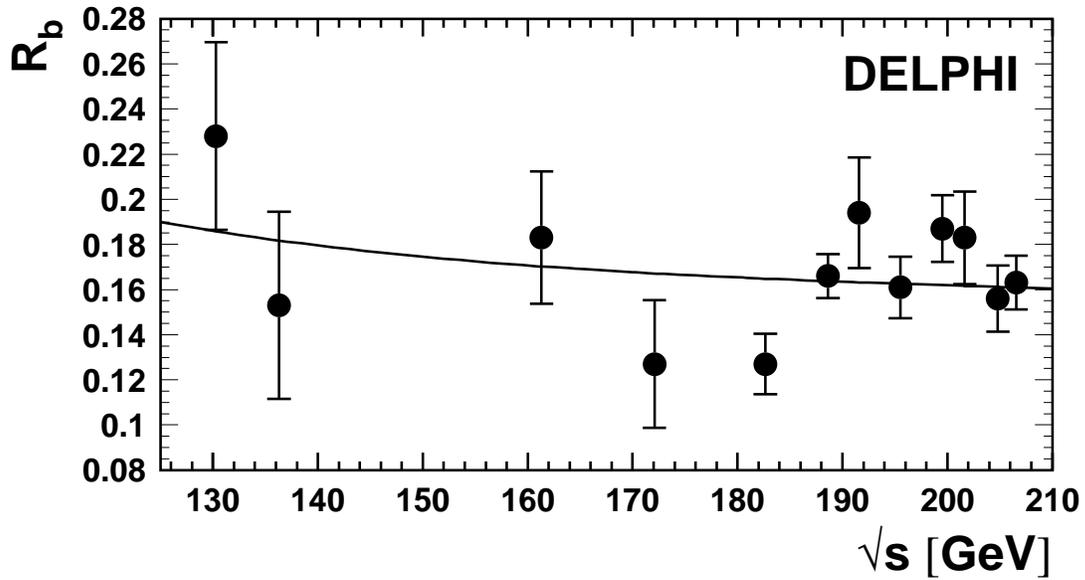,width=0.98\textwidth}
\caption{The measured values (points) of \Rb\ and
the Standard Model predictions (curve)~\cite{ZFIT} plotted
against $\sqrt{s}$. The error bars give the total measurement uncertainties.}
\label{fig:rbres}
\end{center}
\end{figure}

\begin{figure}
\begin{center}
\epsfig{file=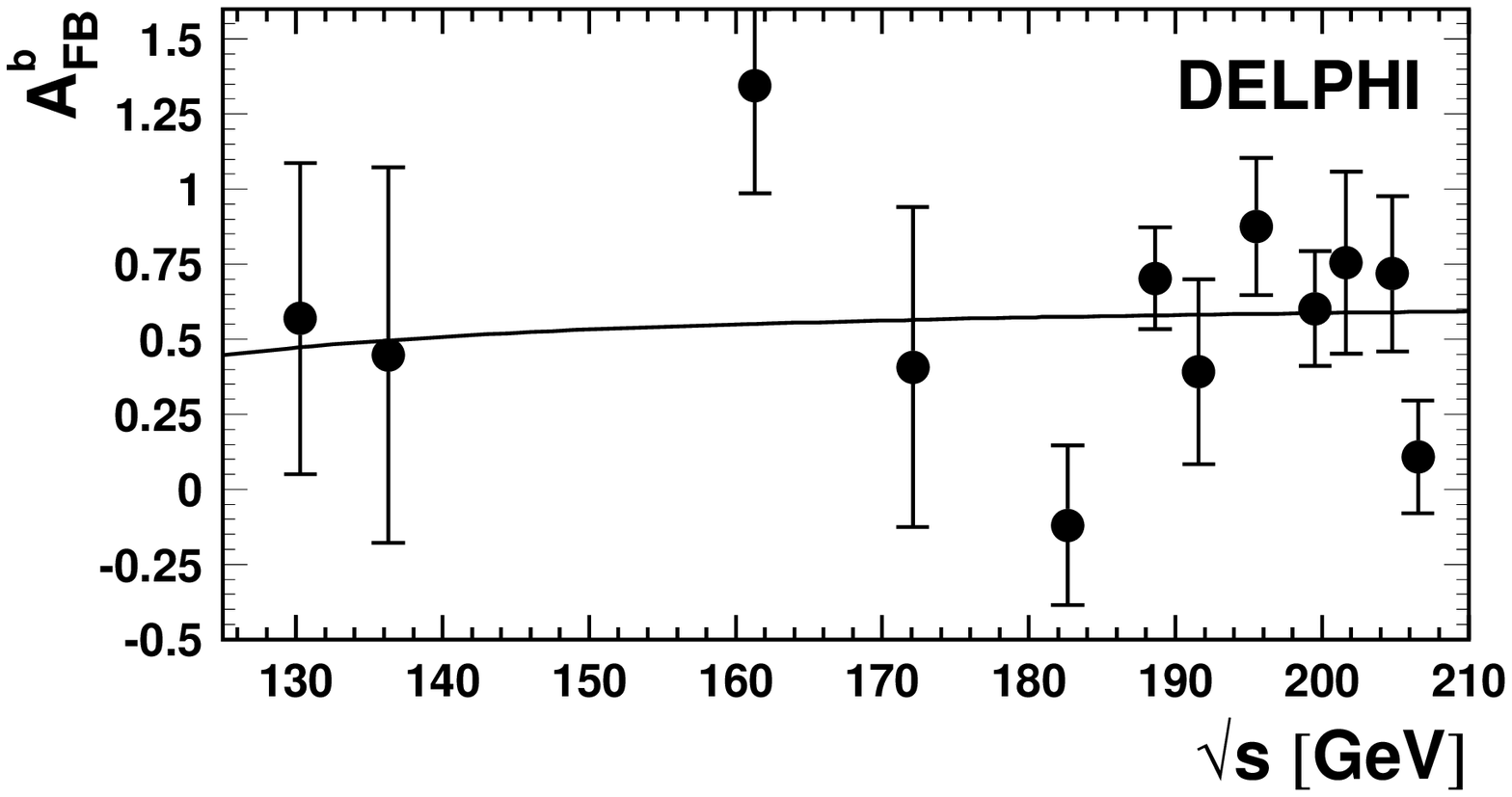,width=0.98\textwidth}
\caption{The measured values (points) of \Ab\ and
the Standard Model prediction (curve)~\cite{ZFIT} plotted
against $\sqrt{s}$. The error bars give the total measurement uncertainties.}
\label{fig:afbres}
\end{center}
\end{figure}

The results for \Rb\ from Sect.~\ref{sec:rbresults} and those for \Ab\ 
from Sect.~\ref{sec:afbresults} have been compared against the 
Standard Model expectations, as calculated by ZFITTER~\cite{ZFIT}
with final state radiation effects included.
The measurements and the expectations are 
shown in Figs.~\ref{fig:rbres} and~\ref{fig:afbres},
for \Rb\ and \Ab\ respectively.   The mean values of the 
differences between the measurements and the Standard Model expectations
have been evaluated using both the statistical and systematic uncertainties,
and taking full account of all
correlations. The results of this computation are presented in Table~\ref{tab:rbafbsm}.
In both cases it can be seen that the measurements agree reasonably well
with the Standard Model.
When all data points are combined, 
the relative precision of the \Rb\ measurements is 3.3\% and the overall
uncertainty on the \Ab\ measurements is 0.083.  These results are the most
precise yet obtained for the two parameters at LEP~2 energies.
%In order to quantify the relative precision of the measurements, 
%the uncertainties found in these tables for the full dataset 
%may be divided by the Standard Model expectations at the
%corresponding mean values of $\sqrt{s}$.
%The relative precision of the \Rb\ measurements is 3.3\%, and that
%of the \Ab\ measurements is 14\%, with negligible correlation between
%the two. 

%\begin{table}
%\begin{center}
%\caption[]{Results of the fit for the mean value 
%of the difference between the measured values 
%of \Rb\ and the Standard Model predictions~\cite{ZFIT}.}
%\label{tab:rbsm}
%\vspace*{0.2cm}
%\begin{tabular}{l|c|c|r}
%Sample   & \multicolumn{1}{|c|}{$<\Rb({\rm Meas - SM})>$} & \multicolumn{1}{c|}{$<\sqrt{s} >$ [GeV]} & 
%\multicolumn{1}{c}{$\chi^2/{\rm ndf}$} \\ \hline
%$\sqrt{s}$ $<$ 190 GeV &  $-0.011 \pm 0.007$ & 181.1 & $9.6/5$ \\
%$\sqrt{s}$ $>$ 190 GeV & \hspace*{0.2cm}  $0.007 \pm 0.007$ & 201.4 & $4.4/5$ \\
%All data             &  $-0.002 \pm 0.005$ & 191.9 & $18.0/11$ \\ 
%\end{tabular}
%\end{center}
%\end{table}

%\begin{table}
%\begin{center}
%\caption[]{Results of the fit for the mean value of the difference between the measured values 
%of \Ab\ and the Standard Model predictions~\cite{ZFIT}.}
%\label{tab:afbsm}
%\vspace*{0.2cm}
%\begin{tabular}{l|c|c|r}
%Sample   & \multicolumn{1}{|c|}{$<\Ab({\rm Meas - SM})>$} & \multicolumn{1}{c|}{$<\sqrt{s} >$ [GeV]} & 
%\multicolumn{1}{c}{$\chi^2/{\rm ndf}$} \\ \hline
%$\sqrt{s}$ $<$ 190 GeV &  $-0.040 \pm 0.119$ & 179.0 & $11.7/5$ \\
%$\sqrt{s}$ $>$ 190 GeV &  $-0.061 \pm 0.104$ & 200.9 & $9.5/5$ \\
%All data             &  $-0.091 \pm 0.081$ & 192.2 & $20.8/11$ \\ 
%\end{tabular}
%\end{center}
%\end{table}

Contact interactions between initial and final state fermionic currents
provide a rather general description of the low energy behaviour of
any new physics process with a characteristic energy scale.  The results of the
\Rb\ and \Ab\ analyses have been compared with a variety of contact
interaction models.  Following reference~\cite{CIREF1} 
the contact interactions are parameterised in the same manner as 
explained in~\cite{DELPHIFF}, in which an effective Lagrangian of
the form:
\begin{equation}
{\cal L}_{eff} = \frac{g^2}{\Lambda^2} \sum_{i,j=L,R} \, \eta_{ij}
\bar{e_i} \gamma_\mu e_i \bar{b_j} \gamma^\mu b_j ,
\end{equation}
is added to the Standard Model Lagrangian.   Here $g^2 /4 \pi$
is taken to be 1 by convention, $\eta_{ij} = \pm 1$ or $0$, $\Lambda$
is the energy scale of the contact interactions, and $e_i$ ($b_j$) are
left or right-handed electron (b-quark) 
spinors.  By assuming different helicity couplings
between the initial-state and final-state currents and either
constructive or destructive interference with the Standard Model
(according to the choice of each $\eta_{ij}$) a set of different
models can be defined from this Lagrangian~\cite{CIREF2}.  The 
values of $\eta_{ij}$ for the models investigated in this
study are given in Table~\ref{tab:cimod}.

\begin{table}
\begin{center}
\caption[]{Results of the fit for the mean value 
of the difference between the measured values 
and the Standard Model predictions~\cite{ZFIT}, for both \Rb\ and \Ab.
The first uncertainty is statistical, and the second uncertainty is systematic.}
\label{tab:rbafbsm}
\vspace*{0.2cm}
\begin{tabular}{l|c|c|rc}
Measurement   & \multicolumn{1}{|c|}{$<({\rm Meas - SM})>$} & \multicolumn{1}{c|}{$<\sqrt{s} >$ [GeV]} & 
\multicolumn{2}{c}{$\chi^2/{\rm ndf}$ (Prob.) } \\ \hline
\Rb\             &  $-0.0016 \pm 0.0044 \pm 0.0031$ & 191.9 & $17.9/11$ & ($8\%$) \\ 
% without QCD error
%\Ab\             &  $-0.091 \pm 0.081$ & 192.2 & $20.8/11$ & ($4\%$) \\ 
% with QCD error
\Ab\             &  $-0.091 \pm 0.072 \pm 0.041$ & 192.2 & $20.8/11$ & ($4\%$) \\ 
\end{tabular}
\end{center}
\end{table}

In fitting for the presence of contact interactions a new parameter
$\epsilon \equiv 1/ \Lambda^2$ is defined, with $\epsilon=0$ being the limit
that there are no new physics contributions.  The region $\epsilon > 0$ 
represents physical values of $1/\Lambda^2$ in models in which there is constructive 
interference with the Standard Model, while the region  $\epsilon < 0$
represents physical values for the equivalent model with destructive
interference.   Least squared fits have been made for the value of 
$\epsilon$ assuming contact interactions from each model listed in
Table~\ref{tab:cimod}.
All \Rb\ and \Ab\ data have been used, taking account of the correlations
between the measurements. In this fit, the \Rb\ results have been
re-expressed as absolute cross-sections, making use of the
\qqbar\ cross-section results found in~\cite{DELPHIFF}.

The results of the contact interaction fits are shown in Table~\ref{tab:cifit}.
The data show no evidence for a non-zero value of $\epsilon$ in any model,
and the table lists the 68\% allowed confidence level
range for the fits to this parameter.  Also shown are 
the corresponding 95\% confidence level lower limits for the contact interaction scale,
allowing for positive ($\Lambda^+$) and negative ($\Lambda^-$) interference
with the Standard Model.  These limits are in the range 2--13 TeV, with
the most stringent for the VV, AA and V0 models.

\begin{table}[htb]
\begin{center}
\caption[]{Choices of $\eta_{ij}$ for different contact interaction models.}
\label{tab:cimod}
\vspace*{0.2cm}
\begin{tabular}{c|r|r|r|r}
Model & $\eta_{LL}$ & $\eta_{RR}$ & $\eta_{LR}$ & $\eta_{RL}$ \\ \hline
LL$^\pm$ & $\pm$ 1 & 0 & 0 & 0 \\
RR$^\pm$ &   0 & $\pm$ 1 & 0 & 0 \\
VV$^\pm$ & $\pm$ 1 & $\pm$ 1 & $\pm$ 1 & $\pm$ 1 \\
AA$^\pm$ & $\pm$ 1 & $\pm$ 1 & $\mp$ 1 & $\mp$ 1 \\
LR$^\pm$ &   0 &  0  & $\pm$ 1 & 0 \\
RL$^\pm$ &   0 &  0  & 0  & $\pm$ 1\\
V0$^\pm$ & $\pm$ 1 & $\pm$ 1 & 0 & 0 \\
A0$^\pm$ &   0 &  0  & $\pm$ 1 & $\pm$ 1 \\
\end{tabular}
\end{center}
\end{table}

\begin{table}
\begin{center}
\caption[]{Limits of contact interactions coupling to \bbbar.  The 68\% C.L.
range is given for $\epsilon$, while 95\% C.L. lower limits are given for $\Lambda^{\pm}$.}
\label{tab:cifit}
\vspace*{0.2cm}
\begin{tabular}{c|r|r|r} 
Model & \multicolumn{1}{|c}{$\epsilon$ (TeV$^{\rm -2}$)} & \multicolumn{1}{|c}{$\Lambda^-$ (TeV)} & 
\multicolumn{1}{|c}{$\Lambda^+$ (TeV)} \\ \hline
%% without QCD correction
%LL & [-0.0020, 0.0098] & 10.1 \hspace*{0.38cm} & 8.4 \hspace*{0.38cm} \\
%RR & [-0.1943, 0.0173] &  2.2 \hspace*{0.38cm}& 5.7 \hspace*{0.38cm}\\
%VV & [-0.0021, 0.0077] & 10.6 \hspace*{0.38cm}& 9.5 \hspace*{0.38cm} \\
%AA & [-0.0012, 0.0061] & 12.8 \hspace*{0.38cm}&10.6 \hspace*{0.38cm} \\
%LR & [-0.1030, 0.0234] &  2.9 \hspace*{0.38cm}& 4.7 \hspace*{0.38cm} \\
%RL & [-0.0162, 0.1688] &  5.8 \hspace*{0.38cm}& 2.4 \hspace*{0.38cm} \\
%V0 & [-0.0014, 0.0070] & 12.0 \hspace*{0.38cm}& 9.9 \hspace*{0.38cm} \\
%A0 & [-0.0163, 0.0632] &  5.3 \hspace*{0.38cm}& 3.7 \hspace*{0.38cm} \\
LL & [-0.0019, 0.0097] & 10.2 \hspace*{0.38cm} & 8.4 \hspace*{0.38cm} \\
RR & [-0.1947, 0.0172] &  2.2 \hspace*{0.38cm}& 5.7 \hspace*{0.38cm}\\
VV & [-0.0021, 0.0076] & 10.6 \hspace*{0.38cm}& 9.5 \hspace*{0.38cm} \\
AA & [-0.0012, 0.0060] & 12.9 \hspace*{0.38cm}&10.7 \hspace*{0.38cm} \\
LR & [-0.1029, 0.0234] &  2.9 \hspace*{0.38cm}& 4.7 \hspace*{0.38cm} \\
RL & [-0.0161, 0.1687] &  5.8 \hspace*{0.38cm}& 2.4 \hspace*{0.38cm} \\
V0 & [-0.0014, 0.0069] & 12.0 \hspace*{0.38cm}& 9.9 \hspace*{0.38cm} \\
A0 & [-0.0163, 0.0630] &  5.3 \hspace*{0.38cm}& 3.7 \hspace*{0.38cm} \\
\end{tabular}
\end{center}
\end{table}

\section{Conclusions}

Analyses of the ratio of the \bbbar\ cross-section to the hadronic cross-section,
\Rb, and the \bbbar\ forward-backward asymmetry, \Ab, have been presented for
non-radiative production, defined as $\sqrt{s'/s} \ge 0.85$, at 12 energy points 
ranging from $\sqrt{s}=$~130.3~GeV to $\sqrt{s}=$~206.6~GeV.   
The relative uncertainties
of all \Rb\ measurements is 3.3\%, and the uncertainty on the mean value of
\Ab\ for all measurements is 0.083, making these results the most 
precise yet obtained for the two parameters at LEP~2 energies.
The results are found to be compatible with those of other 
experiments~\cite{ALEPHBB1,ALEPHBB2,L3BB,OPALBB1,OPALBB2} and are 
consistent with Standard Model expectations.
Limits have been derived on the scales of contact interactions, and are
found to lie in the range 2--13~TeV, depending on the chirality structure of the
new physics contribution.

%\input{acknow.tex}
%         Modified on 04-06-1999 by dimartino
%-------------------------------------------------------------------
\subsection*{Acknowledgements}
\vskip 3 mm
We are greatly indebted to our technical 
collaborators, to the members of the CERN-SL Division for the excellent 
performance of the LEP collider, and to the funding agencies for their
support in building and operating the DELPHI detector.\\
We acknowledge in particular the support of \\
Austrian Federal Ministry of Education, Science and Culture,
GZ 616.364/2-III/2a/98, \\
FNRS--FWO, Flanders Institute to encourage scientific and technological 
research in the industry (IWT) and Belgian Federal Office for Scientific, 
Technical and Cultural affairs (OSTC), Belgium, \\
FINEP, CNPq, CAPES, FUJB and FAPERJ, Brazil, \\
%Czech Ministry of Industry and Trade, GA CR 202/99/1362,\\
%Ministry of Education of the Czech Republic LA134,\\
Ministry of Education of the Czech Republic, project LC527, \\
Academy of Sciences of the Czech Republic, project AV0Z10100502, \\
Commission of the European Communities (DG XII), \\
Direction des Sciences de la Mati$\grave{\mbox{\rm e}}$re, CEA, France, \\
Bundesministerium f$\ddot{\mbox{\rm u}}$r Bildung, Wissenschaft, Forschung 
und Technologie, Germany,\\
General Secretariat for Research and Technology, Greece, \\
National Science Foundation (NWO) and Foundation for Research on Matter (FOM),
The Netherlands, \\
Norwegian Research Council,  \\
State Committee for Scientific Research, Poland, SPUB-M/CERN/PO3/DZ296/2000,
SPUB-M/CERN/PO3/DZ297/2000, 2P03B 104 19 and 2P03B 69 23(2002-2004),\\
FCT - Funda\c{c}\~ao para a Ci\^encia e Tecnologia, Portugal, \\
Vedecka grantova agentura MS SR, Slovakia, Nr. 95/5195/134, \\
Ministry of Science and Technology of the Republic of Slovenia, \\
CICYT, Spain, AEN99-0950 and AEN99-0761,  \\
The Swedish Research Council,      \\
The Science and Technology Facilities Council, UK, \\
%Particle Physics and Astronomy Research Council, UK, \\
Department of Energy, USA, DE-FG02-01ER41155, \\
EEC RTN contract HPRN-CT-00292-2002. \\

%=========================================================================%

\appendix
\section{Correlation Matrices}
\label{sec:cormat}

The correlation matrices for the \Rb\ and \Ab\ results are given in Tables~\ref{tab:corrb} and~\ref{tab:corab}
respectively.   The correlations between \Rb\ and \Ab\ are negligible.

\begin{table}[htb]
\begin{center}
\caption[]{Correlation matrix for \Rb\ results.}
\label{tab:corrb}
\vspace*{0.2cm}
\begin{tabular}{r|rrrrrrrrrrrr}
$\sqrt{s}$ [GeV] & \hspace*{0.15cm} 130 & \hspace*{0.15cm} 136 & \hspace*{0.15cm} 161 &\hspace*{0.15cm} 172 &\hspace*{0.15cm} 183 &\hspace*{0.15cm}  189 & 
\hspace*{0.15cm} 192 & \hspace*{0.15cm} 196 & \hspace*{0.15cm} 200 & \hspace*{0.15cm} 202 & \hspace*{0.15cm} 205 & \hspace*{0.15cm} 207 \\ \hline
130 & 1.00 &  0.01 &  0.01 &  0.01 &  0.02 &  0.01 &  0.02 &  0.03 &  0.03 &  0.02 &  0.03 &  0.04 \\
136 &      &  1.00 &  0.01 &  0.01 &  0.02 &  0.01 &  0.01 &  0.02 &  0.02 &  0.01 &  0.02 &  0.02 \\
161 &      &       &  1.00 &  0.01 &  0.02 &  0.01 &  0.02 &  0.03 &  0.03 &  0.02 &  0.03 &  0.04 \\
172 &      &       &       &  1.00 &  0.02 &  0.01 &  0.01 &  0.02 &  0.03 &  0.02 &  0.02 &  0.03 \\
183 &      &       &       &       &  1.00 &  0.01 &  0.04 &  0.05 &  0.06 &  0.04 &  0.05 &  0.07 \\
189 &      &       &       &       &       &  1.00 &  0.01 &  0.01 &  0.02 &  0.01 &  0.01 &  0.02 \\
192 &      &       &       &       &       &       &  1.00 &  0.06 &  0.06 &  0.04 &  0.04 &  0.06 \\
196 &      &       &       &       &       &       &       &  1.00 &  0.10 &  0.07 &  0.06 &  0.09 \\
200 &      &       &       &       &       &       &       &       &  1.00 &  0.07 &  0.07 &  0.10 \\
202 &      &       &       &       &       &       &       &       &       &  1.00 &  0.05 &  0.07 \\
205 &      &       &       &       &       &       &       &       &       &       &  1.00 &  0.11 \\
207 &      &       &       &       &       &       &       &       &       &       &       &  1.00 \\
\end{tabular}
\end{center}
\end{table}

\begin{table}[htb]
\begin{center}
\caption[]{Correlation matrix for \Ab\ results.}
\label{tab:corab}
\vspace*{0.2cm}
\begin{tabular}{r|rrrrrrrrrrrr}
$\sqrt{s}$ [GeV] & \hspace*{0.15cm} 130 & \hspace*{0.15cm} 136 & \hspace*{0.15cm} 161 &\hspace*{0.15cm} 172 &\hspace*{0.15cm} 183 &\hspace*{0.15cm}  189 & 
\hspace*{0.15cm} 192 & \hspace*{0.15cm} 196 & \hspace*{0.15cm} 200 & \hspace*{0.15cm} 202 & \hspace*{0.15cm} 205 & \hspace*{0.15cm} 207 \\ \hline
130 &  1.00 &  0.04 &  0.04 &  0.02 & -0.06 &  0.07 &  0.02 &  0.04 &  0.04 &  0.03 &  0.04 &  0.05 \\
136 &       &  1.00 &  0.03 &  0.02 & -0.04 &  0.05 &  0.02 &  0.03 &  0.03 &  0.02 &  0.03 &  0.03 \\
161 &       &       &  1.00 &  0.05 & -0.06 &  0.07 &  0.03 &  0.05 &  0.04 &  0.03 &  0.04 &  0.05 \\
172 &       &       &       &  1.00 & -0.04 &  0.04 &  0.02 &  0.03 &  0.03 &  0.02 &  0.03 &  0.03 \\
183 &       &       &       &       &  1.00 & -0.12 & -0.04 & -0.08 & -0.06 & -0.05 & -0.07 & -0.08 \\
189 &       &       &       &       &       &  1.00 &  0.05 &  0.09 &  0.08 &  0.06 &  0.08 &  0.10 \\
192 &       &       &       &       &       &       &  1.00 &  0.04 &  0.04 &  0.03 &  0.03 &  0.04 \\
196 &       &       &       &       &       &       &       &  1.00 &  0.06 &  0.05 &  0.05 &  0.06 \\
200 &       &       &       &       &       &       &       &       &  1.00 &  0.05 &  0.05 &  0.06 \\
202 &       &       &       &       &       &       &       &       &       &  1.00 &  0.04 &  0.04 \\
205 &       &       &       &       &       &       &       &       &       &       &  1.00 &  0.06 \\
207 &       &       &       &       &       &       &       &       &       &       &       &  1.00 \\
\end{tabular}
\end{center}
\end{table}

\end{document}